\documentclass[preprint,3p,onecolumn,12pt]{elsarticle} 
\usepackage{amsmath,amsthm,amscd,color}
\usepackage{amsfonts}
\usepackage{amssymb}
\usepackage{graphicx}
\usepackage{latexsym}
\usepackage{epstopdf}

\usepackage{soul}
\biboptions{sort&compress}
\usepackage{setspace}
\usepackage{xcolor}

\newcommand{\bea}{\begin{eqnarray}}
\newcommand{\eea}{\end{eqnarray}}
\newcommand{\bes}{\begin{subequations}}
\newcommand{\ees}{\end{subequations}}

\newcommand{\ben}{\begin{equation}}
\newcommand{\een}{\end{equation}}
\newcommand{\Del}{\Delta}

\newcommand{\al}{\alpha}
\newcommand{\Lam}{\Lambda}

\newcommand{\vphi}{\varphi}

\newcommand{\sech}{\mbox{ sech}}
\usepackage[colorlinks=true,bookmarks=false,citecolor=blue,urlcolor=blue]{hyperref}

\usepackage{fancyhdr}
\makeatother

\begin{document} 
	\title{General two-component long-wave short-wave resonance interaction system: Non-degenerate vector solitons and their collision dynamics}
	
	\author[bdu1]{S. Stalin \corref{cor}}
	\author[bdu1]{M. Lakshmanan}
	\address[bdu1]{Department of Nonlinear Dynamics, School of Physics, Bharathidasan University, Tiruchirappalli - 620 024, Tamil Nadu, India} 
	
	\cortext[cor]{
		Email: stalin.cnld@gmail.com (S. Stalin)\newline 
		\indent \quad Email: lakshman.cnld@gmail.com (M. Lakshmanan)}
	
	\journal{ Chaos, Solitons \& Fractals}
	
	\setstretch{1.213}
	\begin{abstract}
		In this paper, we demonstrate the emergence of non-degenerate bright solitons and summarize their several interesting features in a completely integrable two-component long-wave-short-wave resonance interaction (LSRI) model with a general form of nonlinearity coefficients. This model describes the nonlinear resonant interaction between a low-frequency long wave and two high-frequency short waves in a one-dimensional physical setting. Through the classical Hirota's bilinear method, we obtain a fully non-degenerate $N$-soliton solution in Gram determinant form for this two-component LSRI model to analyze the nature of non-degenerate vector solitons in detail. Depending on the choice of velocity conditions, the obtained non-degenerate fundamental soliton is classified into two types, namely ($1,1,1$)- and ($1,1,2$)-non-degenerate one solitons. We then show that the basic ($1,1,1$)-non-degenerate soliton exhibits five distinct profile structures, including a novel double-hump, a special flat-top, and a conventional single-hump profile, and ($1,1,2$)-non-degenerate soliton admits two-soliton like oblique collision, a behavior akin to KP line soliton interaction with a short stem structure. A detailed asymptotic analysis is carried out to study the long time behavior of ($1,1,1$)-non-degenerate solitons and it reveals that they undergo both shape-preserving and shape-changing collisions. However, our analysis confirms that the shape changing collision between these solitons become elastic in nature after appropriate shift of time coordinates. Further, we identified that the ($1,1,2$)-non-degenerate solitons also undergo elastic collision. In addition, we have also investigated the formation or suppression of breathing phenomena during collision between a degenerate soliton and a ($1,1,1$)-non-degenerate soliton by considering partially non-degenerate multi-soliton solution. For completeness, we also point out the collision scenario between the completely degenerate solitons. The results presented in this paper are broadly applicable to Bose-Einstein condensates, nonlinear optics, plasma physics, and other closely related fields where the LSRI phenomenon plays a significant role in governing the evolution of vector solitons.\\
		
		\noindent{\it Keywords:} Integrable Long-wave Short-wave Resonance Interaction System, Two-component Yajima-Oikawa system, Non-degenerate Vector Solitons, Degenerate Vector Solitons, Shape Changing Collisions. 
	\end{abstract}
	\maketitle
	
	
	\setstretch{1.3}
	\section{Introduction}
Resonance is a fundamental phenomenon that naturally occurs in both linear and nonlinear dynamical systems when their frequencies satisfy specific conditions \cite{nld-book,sr-book}. This parametric effect is observed in a wide spectrum of physical systems, ranging from simple mechanical oscillators to complex nonlinear systems. Among the different types of resonance phenomena, a particularly intriguing case is the long-wave-short-wave resonance, which emerges from nonlinear interactions between long waves (LWs) and short waves (SWs). This LSRI interaction  emerges when the phase velocity ($v_p=\omega/k$) of the LW matches the group velocity ($v_g=d\omega/dk$) of the SWs. That is, the Zakharov-Benney resonance condition: $v_p=v_g$. This remarkable interaction was first theoretically investigated by V. E. Zakharov in the context of Langmuir wave collapse in plasma \cite{zakharov1}. In his work, he derived a system of coupled equations that would later become known as the generalized Zakharov equations. Meanwhile, the LSRI phenomenon was independently investigated by Benney \cite{benny}, who investigated the interaction dynamics between gravity waves in deep water and short, wind-driven, capillary-gravity waves. Later, Yajima and Oikawa deduced a system, known as the Yajima-Oikawa (YO) system, from the generalized Zakharov equations for describing the unidirectional propagation of Langmuir waves and ion-acoustic waves \cite{oikawa}. Following these initial works, the LSRI phenomenon has received significant attention, leading to numerous theoretical and experimental studies conducted in various physical contexts. Here, we provide a detailed discussion of the developments related to the LSRI phenomenon over the past few decades and  will point out the motivation behind the present study.

The LSRI phenomenon has a wide range of applications, notably in the fields of water wave dynamics, nonlinear optics, Bose-Einstein condensates, plasma physics, and biophysics. To start with, we discuss the appearance of LSRI in water waves. In Ref. \cite{kawahara}, Kawahara {\it et al}. have noticed that the LSRI appears when the short and long capillary gravity waves interact nonlinearly. The LSRI process arises due to the interaction of long and capillary gravity waves in finite-depth water \cite{rede}. Such an interesting LSRI phenomenon was experimentally verified in three-layer fluid flow \cite{rede1}, where the long and short internal waves interact nonlinearly with each other. In addition, it has also been identified in \cite{boyd} when ultralong equatorial Rossby waves interact with short gravity wave groups. The LSRI process has also been recognized in the field of nonlinear optics. In particular, a single-component YO system arises as a reduction of the coupled nonlinear Schr\"odinger equations, which govern the interaction between two optical modes under a small-amplitude limit \cite{kivol}. In negative refractive index media \cite{lsrinim}, the LSRI mechanism arises through a three-wave mixing process, where two degenerate short waves propagate in the negative index branch, while a long wave travels in the positive index branch. The dynamics of a long wave and quasi-resonant two frequency short pulses is governed by the two-component YO system \cite{sazonov}. 

In the context of spinor Bose-Einstein condensates, multicomponent YO system has been derived in \cite {jetp2009}. Through a multiscale expansion approach, higher-dimensional LSRI system has been derived to describe the dynamics of binary disc-shaped BECs \cite{frantz} . The YO system has been obtained to investigate the dynamics of bright–dark soliton complexes in spinor BECs \cite{frantz1}. Multicomponent YO-type systems have also been derived to describe the magnon-phonon interactions \cite{myrza86}. It is important to point out that YO system has been deduced in Ref. \cite{lannig} to study the collision dynamics of mixed vector soliton states in three-component BECs. Besides the above, the LSRI process has also been observed during the nonlinear interaction between an ion-acoustic wave and an electron-plasma wave \cite{nishikawa}. In view of the above, one can visualize 
 the importance styding the LSRI phenomenon and it is clear that the two-component LSRI system considered in the present study possesses substantial physical relevance. 

As it is well known that the model describing the nonlinear interaction between a long wave and a short wave (or multiple short waves) is a completely integrable Hamiltonian system, as demonstrated by Yajima and Oikawa in \cite{oikawa}. So exploring the mathematical aspects of such integrable nonlinear partial differential equations,  including the construction of various nonlinear wave solutions and the analysis of their associated geometrical structures, is another important aspect of the LSRI phenomenon. Therefore, in what follows we emphasize that a wide variety of nonlinear wave solutions have been reported in the literature for the integrable YO-LSRI model and its variants \cite{ma,kanna1,chen1,stalin-lsri,chen2,funakoshi,ohta1,radha,kanna2,kanna3,kanna4,chen3,chow1,crespo1,chow2,crespo2,chen4,kanna5,Gao1,boris-lsri,newell,general-lsri,model-lsri,model-lsri-1,rao}. For the one-dimensional, single-component YO system, both bright and dark soliton solutions were obtained in \cite{ma}. In the case of the $(1+1)$-dimensional multicomponent YO system, fascinating energy-sharing collisions between single-hump bright solitons (we refer these solitons as degenerate vector solitons in our earlier work \cite{stalin-lsri}) were reported \cite{kanna1}. Further, the same multi-component system was shown to admit shape-changing collision scenarios by deriving the mixed bright–dark solitons solutions \cite{chen1}. Novel non-degenerate vector soliton solutions and their several interesting collision dynamics were reported by us in \cite{stalin-lsri} for the two-component LSRI system of YO type.  In contrast, dark solitons in the multicomponent LSRI system exhibit purely elastic collisions, as demonstrated in \cite{chen2}. In the context of two-layer fluid flow, both one- and two-dimensional LSRI systems were derived, with corresponding bright and dark soliton solutions presented in \cite{funakoshi}. A two-component analogue of the two-dimensional LSRI system was formulated by Ohta {\it et al}. by analyzing the nonlinear interactions of dispersive waves across three channels, where soliton solutions in Wronskian form were constructed for the resulting model \cite{ohta1}. This ($2+1$)-dimensional system was further confirmed to be integrable through the Painlev\'e analysis, and dromion solutions were constructed using the Painlev\'e truncation approach \cite{radha}. Notably, one of the present authors (ML) and his collaborators demonstrated energy-sharing collisions of bright solitons in the two-dimensional integrable multicomponent LSRI system by explicitly constructing solutions through the classical Hirota bilinear method \cite{kanna2,kanna3}. The mixed bright-dark soliton solutions and their associated collision dynamics in the $(2+1)$-dimensional multicomponent system have been systematically studied in \cite{kanna4,chen3}, while multi-dark soliton solutions and their elastic collision properties were also examined for the same higher-dimensional LSRI model in \cite{chen2}.   

In addition to the above, rogue waves, a large localized wave structure that appears and vanishes without a trace \cite{akhmediev}, and breathers, which are localized breathing structures, have also been investigated in the LSRI system both in the $(1+1)$- and $(2+1)$-dimensional settings, from single-component to multicomponent systems, revealing interesting dynamical patterns \cite{chow1,crespo1,chow2,crespo2,chen4,kanna5,Gao1,boris-lsri, jrao1}. It should be noted that the variants of the LSRI model, such as the Newell (N) type LSRI system \cite{newell} and the generalized or N-YO type LSRI system \cite{general-lsri}, have been proposed. The various nonlinear wave solutions for these models have been reported in the literature \cite{stalin-nld-lsri,model-lsri,model-lsri-1,rao,jrao2}. Considering all these developments, in this paper, we intend to investigate the recently proposed non-degenerate vector solitons \cite{stalin1,stalin2,stalin3,stalin-review,stalin5,stalin4}, characterized by non-identical wave numbers, and their several novel features in the following two-component LSRI system   
\bea
iS_t^{(l)}+S_{xx}^{(l)}+LS^{(l)}=0,~L_t=\sum_{l=1}^{2}\sigma_l(|S^{(l)}|)_x,~l=1,2. \label{eq.1}
\eea
From the studies reviewed above, this problem remains unexplored in the context of the two-component YO model with a general form of nonlinearity parameters ($\sigma_l$) and so the main objective of the present paper is to investigate this aspect and fill this gap. In Eqs. (\ref{eq.1}), $S^{(l)}$ and $L$, represent the short-waves (SWs) and the long-wave (LW), respectively. The independent variables $x$ and $t$, respectively, denote the partial derivatives with respect to spatial and evolutional coordinates, and $\sigma_l$'s are arbitrary real parameters, which describe the strength of the nonlinearity. 

In order to address this problem, we adopt the classical Hirota bilinear method to derive the $N$-non-degenerate soliton solutions for the considered LSRI model (\ref{eq.1}). These solutions are represented by Gram determinant. We then classify the obtained fundamental non-degenerate soliton solutions based on the velocity conditions into two categories: $(1,1,1)$- and $(1,1,2)$-type non-degenerate solitons. Further, we reveal the collision properties associated with these solitons by performing a detailed asymptotic analysis. In addition to these, we also aim to investigate the collision dynamics between degenerate and non-degenerate solitons by considering the partially non-degenerate two-soliton solution.

The rest of the paper is arranged as follows: In Section 2, we present
the nondegenerate one-soliton solution of the
system (1). This fundamental soliton solution is further classified as ($1,1,1$)- and ($1,1,2$)-non-degenerate fundamental soliton solutions. In this section, we
also discuss the various properties associated with these two basic soliton solutions. In Section 3 we summarize the collision properties of the obtained non-degenerate solitons
with appropriate asymptotic analysis and suitable graphical demonstrations. The degenerate soliton collision-induced novel shape-changing properties of the nondegenerate soliton
is analyzed in Section 4. In Section 5, we summarize the results. In
Appendix A, we provide the $N$-non-degenerate soliton solution in Gram determinant form and asymptotic constants appearing in Section 4 are defined in Appendix B.
	
\section{Non-degenerate fundamental soliton solution and its special features}
We have adopted the classical Hirota's bilinear method to get the exact form of $N$-non-degenerate vector soliton solution of the model (\ref{eq.1}). The corresponding expressions are presented in Appendix A (see Eqs. (\ref{A.2a})-(\ref{A.2d})). To summarize the special features of the fundamental non-degenerate soliton in the present two-component LSRI system (\ref{eq.1}) with arbitrary form of nonlinearity coefficients, in the following we present its explicit form for the $N=1$ case. The form of one-soliton solution is given by
 \bes
 	\bea
 		S^{(1)}(x,t)&=&\frac{\alpha_1^{(1)}e^{\eta_{1}}+e^{\eta_{1}+\xi_{1}+\xi_{1}^{*}+\Del_{1}^{(1)}}}
 		{1+e^{\eta_{1}+\eta_{1}^{*}+R_{1}}+e^{\xi_{1}+\xi_{1}^{*}+R_{2}}+e^{\eta_{1}+\eta_{1}^{*}+\xi_{1}+\xi_{1}^{*}+R_{3}}},\label{2a}\\
 		S^{(2)}(x,t)&=&\frac{\alpha_1^{(2)}e^{\xi_{1}}+e^{\xi_{1}+\eta_{1}+\eta_{1}^{*}+\Del_1^{(2)}}}
 		{1+e^{\eta_{1}+\eta_{1}^{*}+R_{1}}+e^{\xi_{1}+\xi_{1}^{*}+R_{2}}+e^{\eta_{1}+\eta_{1}^{*}+\xi_{1}+\xi_{1}^{*}+R_{3}}},\label{2b}\\
 		L(x,t)&=&2\frac{\partial^2}{\partial x^2}\ln(1+e^{\eta_{1}+\eta_{1}^{*}+R_{1}}+e^{\xi_{1}+\xi_{1}^{*}+R_{2}}+e^{\eta_{1}+\eta_{1}^{*}+\xi_{1}+\xi_{1}^{*}+R_{3}}).\label{2c}
 	\eea

where $\eta_1=k_1x+ik_1^2t$ and  $\xi_1=l_1x+il_1^2t$. The constants involved in the above expressions are defined below:
\bea
&&\hspace{-0.5cm}e^{R_1}=-\frac{i\sigma_1|\al_1^{(1)}|^2}{2(k_1+k_1^*)^2(k_1-k_1^*)},~ e^{R_2}=-\frac{i\sigma_2|\al_1^{(2)}|^2}{2(l_1+l_1^*)^2(l_1-l_1^*)},\nonumber\\ &&\hspace{-0.5cm}e^{\Del_{1}^{(1)}}=-\frac{i\sigma_2\al_1^{(1)}|\al_1^{(2)}|^2(k_1-l_1)}{2(k_1+l_1^*)(l_1-l_1^*)(l_1+l_1^*)^2}, ~ e^{\Del_{1}^{(2)}}=\frac{i\sigma_1\al_1^{(2)}|\al_1^{(1)}|^2(k_1-l_1)}{2(k_1^*+l_1)(k_1-k_1^*)(k_1+k_1^*)^2},\nonumber\\
&&\hspace{-0.5cm}e^{R_3}=-\frac{\sigma_1\sigma_2|\al_1^{(1)}|^2|\al_1^{(2)}|^2|k_1-l_1|^2}{4|k_1+l_1^*|^2(k_1-k_1^*)(l_1-l_1^*)(k_1+k_1^*)^2(l_1+l_1^*)^2}=e^{R_1+R_2+\Lambda},~ e^{\Lambda}=\frac{|k_1-l_1|^2}{|k_1+l_1^*|^2}.\label{2d}
\eea 	 \ees
The above one-soliton solution of the LSRI system (\ref{eq.1}) is characterized by four arbitrary complex parameters: $\alpha_1^{(l)}$, $l=1,2$, $k_1$, and $l_1$ and two system parameters: $\sigma_l$, $l=1,2$.  It is interesting to note that the conventional single-humped `$\sech$' solution can be captured from the general form of one-soliton solution (\ref{2a})-(\ref{2d}) if we set $l_1=k_1$ [so that the wave variable gets restricted as $\xi_1=\xi_{1R}+i\xi_{1I}=\eta_1=\eta_{1R}+i\eta_{1I}=k_{1R}(x-2k_{1I}t)+i(k_{1I}x+(k_{1R}^2-k_{1I}^2)t)$].  Such a restriction yields the standard one-soliton solution of the form \cite{kanna1} 
\bes\bea
\hspace{-0.1cm}S^{(l)}(x,t)=\frac{\alpha_{1}^{(l)}e^{\eta_1}}{1+e^{\eta_1+\eta_1^*+R}}=2A_lk_{1R}\sqrt{k_{1I}}e^{i(\eta_{1I}+\frac{\pi}{2})}\sech (\eta_{1R}+\frac{R}{2}),~A_l=\alpha_1^{(l)}(\sum_{l=1}^{2}\sigma_l|\alpha_{1}^{(l)}|^2)^{-\frac{1}{2}},\label{3a}\\
\hspace{-0.1cm}L(x,t)=2\frac{\partial^2 }{\partial x^2}\ln (1+e^{\eta_1+\eta_1^*+R})=2k_{1R}^2\sech^2(\eta_{1R}+\frac{R}{2}),~R=\ln\frac{-\sum_{l=1}^{2}\sigma_l|\alpha_{1}^{(l)}|^2}{16k_{1R}^2k_{1I}},~ l=1,2.\label{3b}
\eea\ees
The above conventional vector bright soliton is described by only three arbitrary complex parameters,  $\alpha_1^{(l)}$, $l=1,2$, and $k_1$. We remark that one can also obtain the solution (\ref{2a})-(\ref{2d}) from the solution of degenerate two-solitons of the kind (\Ref{3a})-(\ref{3b}) in a non-standard way. The degenerate two-soliton solution can be deduced from  Eqs. (\ref{A.2a})-(\ref{A.2d})) by setting $N=2$ and by restricting $l_1=k_1$ ($\xi_1=\eta_1$) and $l_2=k_2$ ($\xi_2=\eta_2$). By substituting  $\alpha_2^{(1)}=\alpha_1^{(2)}=0$ in such degenerate two-soliton solution and renaming the constants, $\alpha_2^{(2)}$ and $k_2$ as $\alpha_1^{(2)}$ and  as $l_1$, respectively, the solution (\ref{2a})-(\ref{2d}) can be deduced. We have discussed this point in our earlier works in the case of Manakov system \cite{stalin2} and in the case of LSRI system (1) with $\sigma_1=\sigma_2=1$ \cite{stalin-lsri} . However, in principle, it is possible to derive the non-degenerate soliton solution (\ref{2a})-(\ref{2d}) systematically, see Appendix A of the present work and the methodology described in our earlier work \cite{stalin1,stalin2,stalin3,stalin-review,stalin4,stalin5}. The dynamical features of such a new class of vector soliton solution (\ref{2a})-(\ref{2d}) obtained above and the degenerate two soliton solution reported in Ref. \cite{kanna1,stalin-lsri} are entirely different as we will see below. To understand the solution (\ref{2a})-(\ref{2d}) further, we analyse `$|S^{(l)}|^2$' and `$L$' in the limit $|t|\rightarrow \infty$. 

Let us first consider the limit $t\rightarrow -\infty$. In this limit, from the solution (\ref{2a})-(\ref{2d}) under the choice $k_{1I}>l_{1I}$ with $\eta_{1R}\approx 0$ and $\xi_{1R}\rightarrow -\infty$, the expressions for `$|S^{(l)}|^2$' and `$L$' reduce to
\bes\bea
&&|S^{(1)}|^2=\frac{1}{4}e^{\rho_1+\rho_1^*-R_1}\sech^2(\eta_{1R}+\frac{R_1}{2}),~|S^{(2)}|^2=0,\label{4a}\\
&&L=2k_{1R}^2\sech^2(\eta_{1R}+\frac{R_1}{2}),~\eta_{1R}=k_{1R}(x-2k_{1I}t),\label{4b}
\eea\ees
which corresponds to the soliton of the scalar LSRI system (\ref{eq.1}), involving a single SW component $S^{(1)}$ and a LW component with amplitude $\frac{1}{2}e^{\rho_1-\frac{R_1}{2}}$ (for  $S^{(1)}$) and $2k_{1R}^2$ (for $L$), velocity $2k_{1I}$, and central position $\frac{R_1}{2k_{1R}}=\frac{1}{2k_{1R}}\ln\frac{-i\sigma_1|\al_1^{(1)}|^2}{2(k_1+k_1^*)^2(k_1-k_1^*)}$. Note that the soliton appearing in the LW component (see Eq. (\ref{4b})) arises essentially due to the nonlinearity of the system (\ref{eq.1}), specifically, only the soliton associated with the $S^{(1)}$-component contributes to the formation of soliton in the LW component since $|S^{(2)}|^2$ vanishes as specified by Eq. (\ref{4a}).  
Then, by considering the same limit $t\rightarrow -\infty$, we deduce the expressions for `$|S^{(l)}|^2$' and `$L$' when $\xi_{1R}\approx 0$ and $\eta_{1R} \rightarrow +\infty$. The resultant expressions are given by 
\bes\bea
&&|S^{(1)}|^2=0,~|S^{(2)}|^2=\frac{1}{4}e^{\Delta_1^{(2)}+\Delta_1^{(2)*}-(R_1+R_3)}\sech^2(\xi_{1R}+\frac{R_3-R_1}{2}),\label{5a}\\
&&L=2l_{1R}^2\sech^2(\xi_{1R}+\frac{R_3-R_1}{2}),~\xi_{1R}=l_{1R}(x-2l_{1I}t),\label{5b}
\eea\ees
which is again another scalar soliton of the LSRI system (\ref{eq.1}) involving a SW component $S^{(2)}$ and a LW component only with amplitudes $\frac{1}{2}e^{\Delta_1^{(2)}-\frac{(R_1+R_3)}{2}}$ (for  $S^{(2)}$) and $2l_{1R}^2$ (for $L$), velocity $2l_{1I}$, and central position $\frac{R_3-R_1}{2l_{1R}}=\frac{1}{2l_{1R}}\ln\frac{-i\sigma_2|\al_1^{(2)}|^2|k_1-l_1|^2}{2(l_1+l_1^*)^2(l_1-l_1^*)|k_1+l_1^*|^2}$. Here, we observe that the soliton appearing in the SW component $S^{(2)}$ is responsible for the emergence of soliton in the LW component.   

If we now take the limit $t\rightarrow \infty$ with $\eta_{1R}\approx 0$ and $\xi_{1R}\rightarrow \infty$, the expression for `$|S^{(l)}|^2$' and `$L$' are turned out to be
\bes\bea
&&|S^{(1)}|^2=\frac{1}{4}e^{\Delta_1^{(1)}+\Delta_1^{(1)*}-(R_2+R_3)}\sech^2(\eta_{1R}+\frac{R_3-R_2}{2}),~|S^{(2)}|^2=0,\label{6a}\\
&&L=2k_{1R}^2\sech^2(\eta_{1R}+\frac{R_3-R_2}{2}).\label{6b}
\eea\ees
The above expressions show that it is again a scalar soliton of the LSRI system (\ref{eq.1}), involving a single SW component $S^{(1)}$, but now with a different amplitude $\frac{1}{2}e^{\Delta_1^{(1)}-\frac{(R_2+R_3)}{2}}$ and a central position $\frac{R_3-R_2}{2k_{1R}}=\frac{1}{2k_{1R}}\ln\frac{-i\sigma_1|\al_1^{(1)}|^2|k_1-l_1|^2}{2(k_1+k_1^*)^2(k_1-k_1^*)|k_1+l_1^*|^2}$. Similarly, in the limit $t\rightarrow \infty$ and $\xi_{1R}\approx 0$ and $\eta_{1R} \rightarrow -\infty$ we find     
\bes\bea
&&|S^{(1)}|^2=0,~|S^{(2)}|^2=\frac{1}{4}e^{\rho_2+\rho_2^*-R_2}\sech^2(\xi_{1R}+\frac{R_2}{2}),\label{7a}\\
&&L=2l_{1R}^2\sech^2(\xi_{1R}+\frac{R_2}{2}),\label{7b}
\eea\ees
which is again the other soliton of the scalar LSRI system (\ref{eq.1}) involving $S^{(2)}$ SW  component and a LW component with a new amplitude $\frac{1}{2}e^{\rho_2-\frac{R_2}{2}}$ and a new central position $\frac{R_2}{2l_{1R}}=\frac{1}{2l_{1R}}\ln\frac{-i\sigma_2|\al_1^{(2)}|^2}{2(l_1+l_1^*)^2(l_1-l_1^*)}$.    

From the above expressions (4)-(7), we find that changes in amplitudes and position shifts do occur while analysing the asymptotic behaviour of the solution (\ref{2a})-(\ref{2d}). Such a phase shift and variation in amplitude are typical behaviours of multi-soliton solutions of an integrable system, especially in a coupled system. This indicates that solution (\ref{2a})-(\ref{2d}) represents more than just a basic soliton solution or in other words, it is an unconventional form of two-soliton solution. Another notable feature of the solution (\ref{2a})-(\ref{2d}) is that the SW components $S^{(1)}$ and $S^{(2)}$ contain a distinct `$\sech$'-type soliton propagating with velocities $2k_{1I}$ and $2l_{1I}$, respectively. Such two distinct solitons appear in the LW component, through a nonlinear interaction between SW components, exhibiting two-soliton like collision when $k_{1I}\neq l_{1I}$. However, if the solitons of the SW components propagate with equal velocity, specified by $k_{1I}=l_{1I}$, the soliton content gets distributed between the SW components, forming a novel double-hump structure in both the SW components and in the LW component, apart from a special flat-top and a single-hump profiles. Due to the above facts, the  solution (\ref{2a})-(\ref{2d}) can be classified as $(1,1,2)$-non-degenerate soliton solution and $(1,1,1)$-soliton solution if it satisfies the conditions $k_{1I}\neq l_{1I}$ and $k_{1I}=l_{1I}$, respectively. These two basic soliton solutions and their further characterizations are described below.  Note that in our earlier works \cite{stalin1,stalin2,stalin3,stalin4,stalin5,stalin-lsri} and in the present work we have treated the non-degenerate basic soliton solution as the general form of one-soliton solution (for more details see our papers \cite{stalin1,stalin2,stalin-lsri}) and considering the associated  structures emerging from such a solution as a vector one-soliton structure.  We also note that the solution (\ref{2a})-(\ref{2d}) exactly matches with the solution (6a)-(6c) of our paper \cite{stalin-lsri}, where we have shown that Eq. (\ref{eq.1}) with a positive nonlinearity, $\sigma_l=1$, admits non-degenerate soliton solutions. But, in the present work, we summarize the results related to non-degenerate vector solitons for arbitrary choices of the nonlinearity coefficients, including the previously reported case of positive nonlinearity \cite{stalin-lsri}. 
\subsection{(1,1,2) Nondegenerate one-soliton solution}
\begin{figure}
	\centering\includegraphics[width=0.75\linewidth]{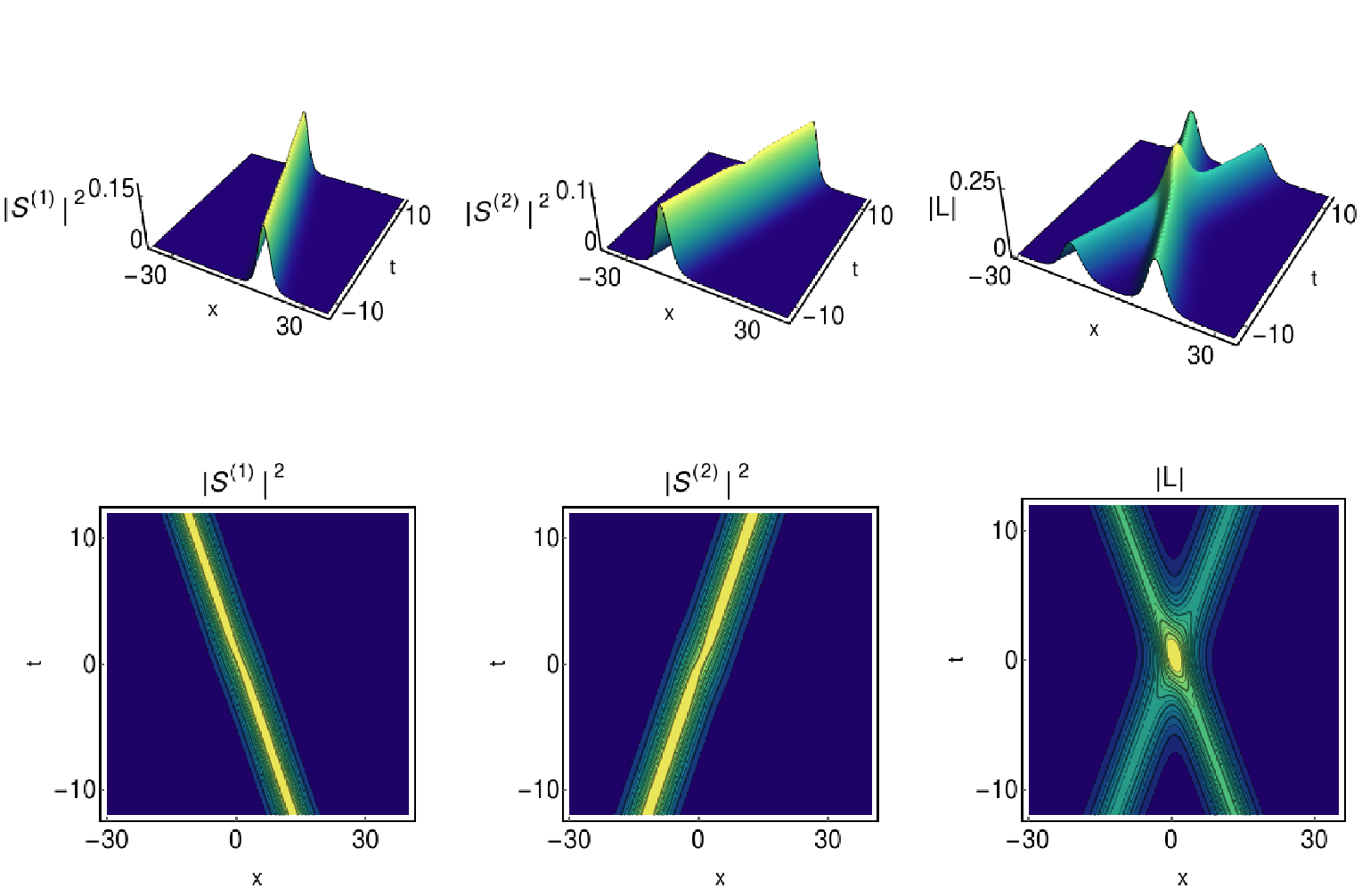} 
	\caption{$(1,1,2)$-nondegenerate fundamental soliton solution: $k_1=0.2-0.5i$, $l_1=0.25+0.5i$, $\alpha_1^{(1)}=0.45+0.5i$, $\alpha_1^{(2)}=0.5+0.5i$, $\sigma_1=-\sigma_2=1$.  }
	\label{fig1}
\end{figure}

 We begin with the choice $k_{1I}\neq l_{1I}$ in Eqs. (\ref{2a})-(\ref{2d}). The resultant solution corresponds to the $(1,1,2)$ non-degenerate one-soliton solution, where each SW component admits a single-hump soliton structure, while the LW component displays a two-soliton-like collision behaviour. The hyperbolic form of this solution reads as
\begin{subequations}
	\begin{eqnarray}
		&&S^{(1)}=\frac{4k_{1R}\sqrt{k_{1I}}e^{i(\eta_{1I}+\theta_1)}\cosh(\xi_{1R}+\phi_1)}{\big[{a_{11}}\cosh(\eta_{1R}+\xi_{1R}+\phi_1+\phi_2+d_{1})+a_{11}^{*-1}\cosh(\eta_{1R}-\xi_{1R}+\phi_2-\phi_1+d_{2})\big]},\label{8a}\\
		&&S^{(2)}=\frac{4l_{1R}\sqrt{l_{1I}}e^{i(\xi_{1I}+\theta_2)}\cosh(\eta_{1R}+\phi_2)}{\big[{a_{12}}\cosh(\eta_{1R}+\xi_{1R}+\phi_1+\phi_2+d_{1})+a_{12}^{*-1}\cosh(\eta_{1R}-\xi_{1R}+\phi_2-\phi_1+d_{2})\big]},\label{8b}\\
		&&L=\frac{4k_{1R}^2\cosh(2\xi_{1R}+2\phi_1+d_3)+4l_{1R}^2\cosh(2\eta_{1R}+2\phi_2+d_4)+d_5}{[e^{\frac{\Lambda}{4}}\cosh(\eta_{1R}+\xi_{1R}+\phi_1+\phi_2+d_{1})+e^{-\frac{\Lambda}{4}}\cosh(\eta_{1R}-\xi_{1R}+\phi_2-\phi_1+d_{2})]^2},\label{8c}
			\end{eqnarray}
\end{subequations}
where\\ $a_{11}=\frac{(k_{1}^{*}-l_{1}^{*})^{\frac{1}{2}}}{(k_{1}^{*}+l_{1})^{\frac{1}{2}}}$,  $a_{12}=\frac{(k_{1}^{*}-l_{1}^{*})^{\frac{1}{2}}}{(k_{1}+l_{1}^{*})^{\frac{1}{2}}}$,   $d_1=\frac{1}{2}\log\frac{(k_1^*-l_1^*)}{(l_1-k_1)}$, $d_2=\frac{1}{2}\log\frac{(k_1-l_1)(k_1^*+l_1)}{(l_1-k_1)(k_1+l_1^*)}$, $d_3=\frac{1}{2}\log\frac{(k_1^*-l_1^*)(k_1+l_1^*)}{(k_1^*+l_1)(k_1-l_1)}$, $d_4=\frac{1}{2}\log\frac{(k_1^*-l_1^*)(k_1^*+l_1)}{(k_1+l_1^*)(l_1-k_1)}$,
$d_5=2((k_{1R}-l_{1R})^2e^{-\frac{\Lambda}{2}}+(k_{1R}+l_{1R})^2e^{\frac{\Lambda}{2}})$, 
 $\eta_{1R}=k_{1R}(x-2k_{1I}t)$, $\eta_{1I}=k_{1I}x+(k_{1R}^2-k_{1I}^2)t$, $\xi_{1R}=l_{1R}(x-2l_{1I}t)$, $\xi_{1I}=l_{1I}x+(l_{1R}^2-l_{1I}^2)t$,
$e^{i\theta_1}=i[\alpha_{1}^{(1)}/\sigma_1\alpha_{1}^{(1)*}]^{1/2}$, $e^{i\theta_2}=i[\alpha_{1}^{(2)}/\sigma_2\alpha_{1}^{(2)*}]^{1/2}$, $\Lambda=\ln\frac{|k_1-l_1|^2}{|k_1+l_1^*|^2}$,  $\phi_1=\frac{1}{2}\log\frac{\sigma_2(l_1-k_1)|\alpha_{1}^{(2)}|^2}{16l_{1R}^2l_{1I}(k_1+l_1^*)}$, and $\phi_2=\frac{1}{2}\log\frac{\sigma_1(k_1-l_1)|\alpha_{1}^{(1)}|^2}{16k_{1R}^2k_{1I}(k_1^*+l_1)}$. Here, $k_{1R}$, $l_{1R}$, $k_{1I}$ and $l_{1I}$ denote the real and imaginary parts of $k_1$ and $l_1$, respectively.  The four arbitrary complex parameters, $\al_1^{(l)}$, $l=1,2$, $k_1$ and $l_1$, determine the  structure of the $(1,1,2)$-non-degenerate fundamental soliton solution of the two-component LSRI system (\ref{eq.1}). The constants, $4 k_{1R} \sqrt{k_{1I}}$ and $4 l_{1R} \sqrt{l_{1I}}$, respectively,  appearing in Eqs. (\ref{8a})-(\ref{8c}) define the amplitudes of the solitons in the SW components. In the LW component, real parts of $k_1$ and $l_1$ describe the amplitudes of solitons.  The velocities of the SW component solitons are $2k_{1I}$ and $2l_{1I}$ whereas the velocities of solitons in the LW component are defined by the imaginary parts of wave numbers $k_1$ and $l_1$ since, as we explained earlier, the SW component solitons appear in the LW component specified by the nonlinearity of the system (\ref{eq.1}). An interesting feature of the solution (\ref{8a})-(\ref{8c}) is the explicit dependence of the amplitudes of the SW component  on the soliton velocity. As a result, the taller non-degenerate soliton propagates at a higher speed, resembling the behaviour of KdV solitons: $u(x,t)=\frac{c}{2}\sech^2\frac{\sqrt{c}}{2}(x-ct)$, where $c$ is the velocity of the basic KdV soliton. The presence of amplitude-dependent velocity is a distinctive property of long-wave short-wave resonance interaction. Another interesting feature of the ($1,1,2$)-soliton solution is the appearance of two-soliton like elastic collision in the LW component whereas in each of the SW components a single-humped soliton structure emerges and propagates in respective directions. Such a possibility can be observed from Fig. \ref{fig1}, where this characteristic is demonstrated for the mixed nonlinearity case: $\sigma_1=-\sigma_2=1$.

\begin{figure}
	\centering\includegraphics[width=0.7\linewidth]{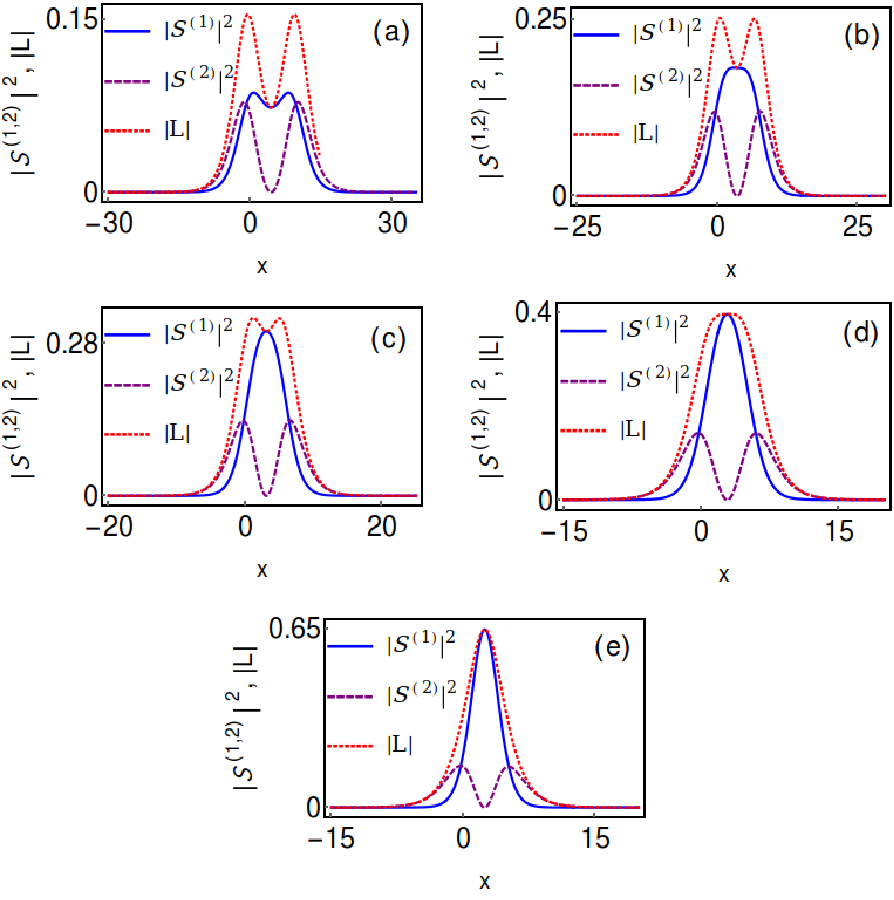} 
	\caption{Five distinct profile structures of the obtained $(1,1,1)$-nondegenerate fundamental soliton solution: Panel (a): $k_1=0.315+0.5i$, $l_1=0.25+0.5i$, $\alpha_1^{(1)}=0.45+0.5i$, and $\alpha_1^{(2)}=0.5+0.5i$. Panel (b): $k_1=0.425+0.5i$, $l_1=0.3+0.5i$, $\alpha_1^{(1)}=0.45+0.45i$, and $\alpha_1^{(2)}=0.43+0.55i$. Panel (c): $k_1=0.5+0.5i$, $l_1=0.315+0.5i$, $\alpha_1^{(1)}=0.45+0.45i$, and $\alpha_1^{(2)}=0.5+0.5i$. Panel (d): $k_1=0.545+0.5i$, $l_1=0.315+0.5i$, $\alpha_1^{(1)}=0.45+0.45i$, and $\alpha_1^{(2)}=0.5+0.5i$. Panel (e): $k_1=0.65+0.5i$, $l_1=0.315+0.5i$, $\alpha_1^{(1)}=0.45+0.45i$, and $\alpha_1^{(2)}=0.5+0.5i$.}
	\label{fig2}
\end{figure}

The solution (\ref{8a})-(\ref{8c}) admits both regular and singular solitons. The singularity property of this solution is mainly determined by the constants: $e^{R_1}$, $e^{R_2}$ and $e^{R_3}$ (see Eq. (\ref{2d})). In the following, the condition for regular soliton solution is described for all the nonlinearity cases.\\
(i) Positive nonlinearity ($\sigma_l>0$): In this case, the regular solution arises when $k_{1I}<0$ and $l_{1I}<0$. Under this condition, the constants, $e^{R_n}$, $n=1,2,3$, are strictly positive. This constraint ensures that solitons in the LSRI model (\ref{eq.1}), with positive nonlinearity, to propagate in the $-x$-direction, as illustrated in Fig. 3 of our earlier paper Ref. \cite{stalin-lsri}.  \\
(ii) Negative nonlinearity ($\sigma_l<0$): For the choice $\sigma_l<0$, obtaining a regular solution requires $e^{R_n}$ ($n=1,2,3$), to be strictly positive, which in turn necessitates $k_{1I}, l_{1I}>0$. As a result, the solitons propagate in the opposite direction compared to the previous case. That is, solitons in all the components travel
to the $+x$-direction. For brevity, we do not illustrate this scenario here. \\
(iii) Mixed nonlinearity ($\sigma_1>0$ \& $\sigma_2<0$ or $\sigma_1<0$ \& $\sigma_2>0$): When considering a mixed choice of nonlinearity strengths to obtain a non-singular solution, the constants $e^{R_n}$ ($n=1,2,3$), which appear in the denominators of Eqs. (\ref{2a})-(\ref{2c}), must be positive definite. For instance, if $\sigma_1>0$ and $\sigma_2<0$, a regular solution requires $k_{1I}<0$ and $l_{1I}>0$. On the other hand, if $\sigma_1<0$ and $\sigma_2>0$, the condition is reversed, that is, $k_{1I}>0$ and $l_{1I}<0$. This implies that the solitons in the two SW components propagate in opposite directions as shown in Fig. \ref{fig1}. However, in the LW component, these solitons appear and exhibit a two-soliton-like oblique collision. This collision scenario closely resembles a resonant interaction between two Kadomtsev-Petviashvili (KP) line-solitons with a short stem structure \cite{kp-stem,biondini-stem,he-stem}. A summary of the propagation properties of the ($1,1,2$)-fundamental non-degenerate soliton solution (\ref{8a})-(\ref{8c}) is provided in Table 1.

\begin{table}
	\centering
	\caption{Propagation properties of (1,1,2)-fundamental non-degenerate soliton }
	\begin{tabular}{|p{3.4cm}|p{3.98cm}|p{3.9cm}|p{3.8cm}|}
		\hline\hline
		Soliton Characteristics & Positive nonlinearity: $\sigma_l>0$, $l=1,2$  & Negative nonlinearity: $\sigma_l<0$, $l=1,2$   & Mixed nonlinearity: $\sigma_1>0$, $\sigma_2<0$ (or $\sigma_1<0$, $\sigma_2>0$) \\
		\hline
		Condition for regular soliton &  $k_{1I},l_{1I}<0$   & $k_{1I},l_{1I}>0$  &   $k_{1I}<0$, $l_{1I}>0$ (or $k_{1I}>0$, $l_{1I}<0$) \\
		\hline
		$S^{(1)}$ component: Amplitude ($A_{1}$) \& Velocity ($v_{1}$) &  $A_1=4k_{1R}\sqrt{|k_{1I}|}$; $v_1=-2|k_{1I}|$ -- soliton moving in $-x$-direction & $A_1=4k_{1R}\sqrt{k_{1I}}$; $v_1=2k_{1I}$ -- soliton moving in $+x$-direction  &   $A_1$ \& $v_1$: $4k_{1R}\sqrt{|k_{1I}|}$ \& $-2|k_{1I}|$ (or $4k_{1R}\sqrt{k_{1I}}$ \& $2k_{1I}$) -- soliton moving in $-x$ (or $+x$)-direction\\
		\hline
		$S^{(2)}$ component: Amplitude ($A_{2}$) \& Velocity ($v_{2}$)&  $A_2=4l_{1R}\sqrt{|l_{1I}|}$; $v_2=-2|l_{1I}|$  -- soliton moving in $-x$-direction& $A_2=4l_{1R}\sqrt{l_{1I}}$; $v_2=2l_{1I}$ soliton moving in $+x$-direction &   $A_2$ \& $v_2$: $4l_{1R}\sqrt{|l_{1I}|}$ \& $-2|l_{1I}|$ (or $4l_{1R}\sqrt{l_{1I}}$ \& $2l_{1I}$) -- soliton moving in $-x$ (or $+x$)-direction \\
		\hline
		LW component ($L$) &  Amplitude and velocity: R.P and I.P of $k_1$ and $l_1$, respectively. 2-soliton like overtaking-collision in the $-x$-direction.&  Amplitude and velocity: R.P and I.P of $k_1$ and $l_1$, respectively. 2-soliton like overtaking-collision in the $+x$-direction. &     Amplitude and velocity: R.P and I.P of $k_1$ and $l_1$, respectively. 2-soliton like head-on collision.\\
		\hline\hline
	\end{tabular}
	\label{table-1}
\end{table}	
	
\subsection{$(1,1,1)$ Nondegenerate one-soliton solution}
 Now, we consider the other choice $k_{1I}=l_{1I}$ in the solution (\ref{8a})-(\ref{8c}). The final outcome is nothing  but a $(1,1,1)$-non-degenerate one-soliton solution, in which all the SW components and the LW component exhibit a combination of double-hump, flat-top, and single-hump profiles. The soliton structures emerge from such a solution propagate in all the components with identical/equal velocities. To represent such soliton structures mathematically we impose $k_{1I}=l_{1I}$ in the $(1,1,2)$-solution (\ref{8a})-(\ref{8c}). This process yields the following hyperbolic form of ($1,1,1$)-soliton solution, which reads as
\begin{subequations}
	\begin{eqnarray}
		&&S^{(1)}=\frac{4k_{1R}\sqrt{k_{1I}}e^{i(\eta_{1I}+\theta_1)}\cosh(\xi_{1R}+\phi_{1R})}{\big[{a_{11}}\cosh(\eta_{1R}+\xi_{1R}+\phi_{1R}+\phi_{2R}+\frac{i\pi}{2})+a_{11}^{-1}\cosh(\eta_{1R}-\xi_{1R}+\phi_{2R}-\phi_{1R}+\frac{i\pi}{2})\big]},~~~~~\label{9a}\\
		&&S^{(2)}=\frac{4l_{1R}\sqrt{k_{1I}}e^{i(\xi_{1I}+\theta_2)}\cosh(\eta_{1R}+\phi_{2R})}{\big[a_{11}\cosh(\eta_{1R}+\xi_{1R}+\phi_{1R}+\phi_{2R}+\frac{i\pi}{2})+a_{11}^{-1}\cosh(\eta_{1R}-\xi_{1R}+\phi_{2R}-\phi_{1R}+\frac{i\pi}{2})\big]},\label{9b}\\
		&&L=\frac{4k_{1R}^2\cosh(2\xi_{1R}+2\phi_{1R})+4l_{1R}^2\cosh(2\eta_{1R}+2\phi_{2R}+\frac{i\pi}{2})+4(k_{1R}^2-l_{1R}^2)}{[a_{11}\cosh(\eta_{1R}+\xi_{1R}+\phi_{1R}+\phi_{2R}+\frac{i\pi}{2})+a_{11}^{-1}\cosh(\eta_{1R}-\xi_{1R}+\phi_{2R}-\phi_{1R}+\frac{i\pi}{2})]^2},\label{9c}
	\end{eqnarray}
\end{subequations}
where\\ $a_{11}=\frac{(k_{1R}-l_{1R})^{\frac{1}{2}}}{(k_{1R}+l_{1R})^{\frac{1}{2}}}$,  
$\eta_{1R}=k_{1R}(x-2k_{1I}t)$, $\eta_{1I}=k_{1I}x+(k_{1R}^2-k_{1I}^2)t$, $\xi_{1R}=l_{1R}(x-2k_{1I}t)$, $\xi_{1I}=k_{1I}x+(l_{1R}^2-k_{1I}^2)t$,
$e^{i\theta_1}=i[\alpha_{1}^{(1)}/\sigma_1\alpha_{1}^{(1)*}]^{1/2}$, $e^{i\theta_2}=i[\alpha_{1}^{(2)}/\sigma_2\alpha_{1}^{(2)*}]^{1/2}$, $\phi_{1R}=\frac{1}{2}\log\frac{\sigma_2(l_{1R}-k_{1R})|\alpha_{1}^{(2)}|^2}{16l_{1R}^2l_{1I}(k_{1R}+l_{1R})}$, and $\phi_{2R}=\frac{1}{2}\log\frac{\sigma_1(k_{1R}-l_{1R})|\alpha_{1}^{(1)}|^2}{16k_{1R}^2k_{1I}(k_{1R}+l_{1R})}$. From Eqs. (\ref{8a})-(\ref{8c}), one is able to identify a relation,
\begin{equation}
	L=-\frac{1}{2k_{1I}}(\sigma_1|S^{(1)}|^2+\sigma_2|S^{(2)}|^2).\label{10}
\end{equation}
between the LW component and SW components. This relation again shows that the formation of soliton structure in the LW component is essentially due to the linear superposition of intensities of the SW components. The two arbitrary complex parameters, $\al_1^{(l)}$'s, $l=1,2$, and two wave numbers, $k_1$ and $l_1$, having equal imaginary parts, determine the  structure of the ($1,1,1$) non-degenerate fundamental soliton solution (\ref{8a})-(\ref{8c}) of the two-component LSRI system (\ref{eq.1}). 

The solution (\ref{9a})-(\ref{9c}) admits five sets of profile structures, including a double-hump, a special flat-top and a single-hump profiles for $k_{1R}>l_{1R}$ (or $k_{1R}<l_{1R}$). The combination of such profiles which emerge from the ($1,1,1$)-non-degenerate one-soliton solution (\ref{9a})-(\ref{9c}) are defined as follows: (i) double-hump in all components, (ii) a flat-top in the $S^{(1)}$ component and double-hump in the remaining components, (iii) a single-hump in the $S^{(1)}$ component and double-hump in the remaining components, (iv) a double-hump in the $S^{(2)}$ component and a single-hump in the $S^{(1)}$ component and a flat-top in the $L$ component, and (v) a double-hump in the $S^{(2)}$ component and single-hump profiles in both the $S^{(1)}$ and $L$ components. These five distinct set of profiles are displayed in Fig. \ref{fig2} for the parameter values defined therein. 
\begin{table}
	\centering
	\caption{(1,1,1)-fundamental non-degenerate soliton properties}
	\begin{tabular}{|p{3.5cm}|p{3.9cm}|p{3.94cm}|p{3.8cm}|}
		\hline\hline
		Soliton Characteristics & Positive nonlinearity: $\sigma_l>0$, $l=1,2$  & Negative nonlinearity: $\sigma_l<0$, $l=1,2$   & Mixed nonlinearity: $\sigma_1>0$, $\sigma_2<0$ (or $\sigma_1<0$, $\sigma_2>0$) \\
		\hline
		Condition for regular soliton &  $k_{1I}<0$   & $k_{1I}>0$  &   always singular \\
		\hline
		SW components $S^{(1)}$ \& $S^{(2)}$: Amplitudes ($A_{1}$ \& $A_{2}$) \& Velocity ($v_{1}$) &  $S^{(1)}\rightarrow $ $A_1=4k_{1R}\sqrt{|k_{1I}|}$; $S^{(2)}\rightarrow $ $A_2=4l_{1R}\sqrt{|k_{1I}|}$; $v_1=-2|k_{1I}|$ -- soliton moving in $-x$-direction  & $S^{(1)}\rightarrow $ $A_1=4k_{1R}\sqrt{k_{1I}}$; $S^{(2)}\rightarrow $ $A_2=4l_{1R}\sqrt{k_{1I}}$; $v_1=2k_{1I}$ -- soliton moving in $+x$-direction   &  singular soliton \\
		\hline
		LW component &  Amplitude and velocity: R.P and I.P of $k_1$ and $l_1$, respectively. &  Amplitude and velocity: R.P and I.P of $k_1$ and $l_1$, respectively. &      singular soliton \\
		\hline\hline
	\end{tabular}
	\label{table-2}
\end{table}

The solution (\ref{9a})-(\ref{9c}) also admits both regular and singular solitons. Here, also the singularity property of this solution is mainly determined by the constants: $e^{R_1}$, $e^{R_2}$ and $e^{R_3}$. The condition for regular soliton solution is described for all the nonlinearity cases as follows.\\
(i) Positive nonlinearity ($\sigma_l>0$): In this case, the regular solution arises when $k_{1I}<0$. Under this condition, the constants, $e^{R_n}$, $n=1,2,3$, should be strictly positive. This requirement guarantees that solitons in the LSRI model (\ref{eq.1}) with positive nonlinearity propagates in the $-x$-direction \cite{stalin-lsri}. \\ 
(ii) Negative nonlinearity: For the choice $\sigma_l<0$, to obtain a regular solution, the quantity $e^{R_n}$ ($n=1,2,3$), should be strictly positive, which requires $k_{1I}>0$. Therefore, the solitons propagate to the $+x$-direction. \\
(iii) Mixed nonlinearity: When a mixed choice of nonlinearity is considered, that is $\sigma_1>0$ \& $\sigma_2<0$ or $\sigma_1<0$ \& $\sigma_2>0$, the ($1,1,1$) non-degenerate soliton solution (\ref{9a})-(\ref{9c}) always admits singular soliton. We have summarized the propagation properties of the ($1,1,1$)-fundamental non-degenerate soliton solution (\ref{9a})-(\ref{9c}) in Table 2.      

\begin{figure}[h]
	\centering
	\includegraphics[width=0.95\linewidth]{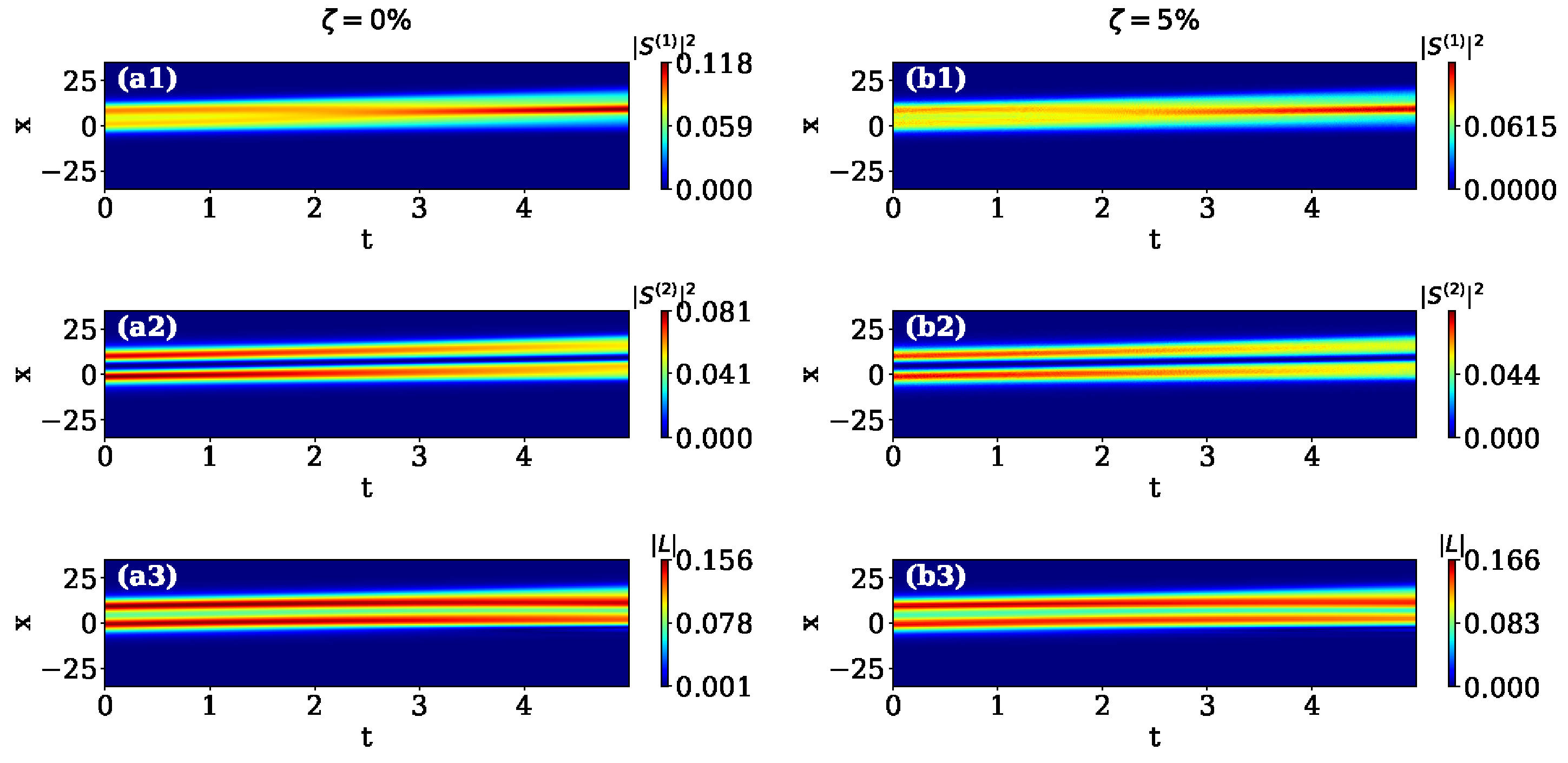}
	\caption{Panels (a1)-(a3) illustrate the propagation of the double-hump profile of the non-degenerate soliton in the absence of perturbation, while panels (b1)-(b3) demonstrate the stability of the same double-hump soliton under 5\% random noise perturbation.} 
	\label{stability}
\end{figure}

We have also verified the stability of the obtained $(1,1,1)$-non-degenerate one soliton solution (\ref{9a})-(\ref{9c}) by adopting the split-step Fourier method numerically by introducing random noise as a perturbation in the interval $[-1,1]$. For this analysis, we have considered the double-hump profiles in all three components as the initial condition with spatial and temporal step sizes $\Delta x= 0.1$ and $\Delta t=0.001$, respectively. The computational domain is chosen as $[-30,30]$ for $x$ and $[0,5]$ for $t$. The numerical results are presented in Fig. \ref{stability}, which show that the double-hump profile of the non-degenerate soliton persists even under a $5\%$ of random perturbation. This confirms that the obtained non-degenerate solitons are stable against perturbations.     

\section{Collision dynamics of non-degenerate solitons}\label{collisions} 
In this section, we examine the collision properties of the non-degenerate solitons of system (\ref{eq.1}) and reveal the underlying mechanism responsible for their intriguing collision dynamics. As explained earlier, the fundamental non-degenerate soliton can exist in two forms: the ($1,1,1$)-non-degenerate one-soliton and the ($1,1,2$)-non-degenerate one-soliton. Accordingly, it is important to analyze the following three collision scenarios: (i) collision between two ($1,1,1$)-non-degenerate solitons, (ii) interaction between two ($1,1,2$)-non-degenerate solitons, and (iii) collision between ($1,1,1$)- and  ($1,1,2$)-non-degenerate solitons. To investigate these collision scenarios, one has to analyse the non-degenerate two soliton solution ($N=2$ in Eqs. (\ref{A.2a})-(\ref{A.2d})) in the long time limits $t\rightarrow\pm\infty$ with appropriate choices of velocity restrictions. We have carried out necessary asymptotic analysis to systematically examine the interaction between ($1,1,1$)-non-degenerate solitons.  In principle, a similar analysis can be performed for the other cases as well. The corresponding detailed mathematical derivations are not presented here to maintain brevity. However, we have illustrated these cases graphically by providing appropriate explanations. We note that the aforementioned collision scenarios have been investigated in Ref. \cite{stalin-lsri} by considering the system (\ref{eq.1})  with $\sigma_l=1$, $l=1,2$. Here, the results are presented for system (\ref{eq.1}) with negative nonlinear coefficients ($\sigma_l=-1$, $l=1,2$). 

	\subsection{Collision between $(1,1,1)$-non-degenerate solitons}
As we have mentioned earlier, to study the collision among two ($1,1,1$)-non-degenerate solitons, it is necessary to choose $k_{jI}=l_{jI}$, for $j=1,2$, in the two-soliton solution, which can be deduced from Eqs. (\ref{A.2a})-(\ref{A.2d}) by restricting to $N=2$. Then, we also further consider the parametric choices, $k_{jI}=l_{jI}>0$, $k_{jR}, l_{jR}>0$, $j=1,2$, and $k_{1I}<k_{2I}$ (or $l_{1I}<l_{2I}$). Note that the singularity condition: $k_{jI}=l_{jI}>0$, for negative nonlinearity coefficients, enforces the ($1,1,1$)-non-degenerate solitons to propagate in the $+x$-direction so that an overtaking collision only occurs in the same direction. Thus, head-on collision is not possible between the ($1,1,1$)-non-degenerate solitons in the case of negative nonlinearity.  To deduce the asymptotic forms of these solitons in the long time regimes, we incorporate the asymptotic behaviour of the wave variables, $\eta_{jR}=k_{jR}(x-2k_{jI}t)$ and  $\xi_{jR}=l_{jR}(x-2k_{jI}t)$, $j=1,2$, in the corresponding two-soliton solution. By doing so, we have deduced the following asymptotic forms of ($1,1,1$)-non-degenerate solitons. 
\\
\underline{(a) Before collision}: $t\rightarrow -\infty$\\
\underline{Soliton 1}: $\eta_{1R}$, $\xi_{1R}\simeq 0$, $\eta_{2R}$, $\xi_{2R}\rightarrow+ \infty$\\
The asymptotic forms of $S^{(l)}$, $l=1,2$ and $L$ are deduced for soliton 1 and they read as 
\begin{eqnarray}
	&&S^{(1)}\simeq \frac{4k_{1R}\sqrt{k_{1I}}e^{i(\eta_{1I}+\theta_1^{1-}+\frac{\pi}{2})}\cosh(\xi_{1R}+\phi_1^-)}{\big[{a_{11}}\cosh(\eta_{1R}+\xi_{1R}+\phi_1^-+\phi_2^-+c_1)+a_{11}^{*-1}\cosh(\eta_{1R}-\xi_{1R}+\phi_2^--\phi_1^-+c_2)\big]},\nonumber\\
	&&S^{(2)}\simeq\frac{4l_{1R}\sqrt{k_{1I}}e^{i(\xi_{1I}+\theta_2^{1-}+\frac{\pi}{2})}\cosh(\eta_{1R}+\phi_2^-)}{\big[a_{12}\cosh(\eta_{1R}+\xi_{1R}+\phi_1^-+\phi_2^-+c_1)+a_{12}^{*-1}\cosh(\eta_{1R}-\xi_{1R}+\phi_2^--\phi_1^-+c_2)\big]},\nonumber\\
	&&L\simeq\frac{4}{f^2}\bigg((k_{1R}^2-l_{1R}^2)+l_{1R}^2\cosh(2\eta_{1R}+2\phi_2^-+c_3)+k_{1R}^2\cosh(2\xi_{1R}+2\phi_1^-+c_4)\bigg),\nonumber\\
	&&f=b_1\cosh(\eta_{1R}+\xi_{1R}+\phi_1^-+\phi_2^-+c_1)+b_1^{-1}\cosh(\eta_{1R}-\xi_{1R}+\phi_2^--\phi_1^-+c_2). \label{11}
\end{eqnarray}
Here,  $\phi_{1}^{-}=\frac{1}{2}\ln \frac{i(k_1-l_1)|k_2-l_1|^2|l_1-l_2|^4|l_1+l_2|^2|\alpha_{1}^{(2)}|^2}{2(l_1-l_1^*)(l_1+l_1^*)^2|k_2+l_1^*|^2(k_1+l_1^*)|l_1-l_2^*|^2|l_1+l_2^*|^4}$, $a_{11}=\frac{(k_{1}^{*}-l_{1}^{*})^{\frac{1}{2}}}{(k_{1}^{*}+l_{1})^{\frac{1}{2}}}$,  $a_{12}=\frac{(k_{1}^{*}-l_{1}^{*})^{\frac{1}{2}}}{(k_{1}+l_{1}^{*})^{\frac{1}{2}}}$,  \\
$e^{i\theta_1^{1-}}=\frac{(k_{1}-k_{2})(k_{1}+k_{2})^{\frac{1}{2}}(k_{1}^*+k_{2})(k_{1}-l_{2})^{\frac{1}{2}}(k_{1}^*-k_{2})^{\frac{1}{2}}(k_{1}^{*}+l_{2})^{\frac{1}{2}}\sqrt{\alpha_1^{(1)}}}{(k_{1}^{*}-k_{2}^{*})(k_{1}^*+k_{2}^*)^{\frac{1}{2}}(k_{1}+k_{2}^*)(k_{1}^{*}-l_{2}^{*})^{\frac{1}{2}}(k_{1}-k_{2}^*)^{\frac{1}{2}}(k_{1}+l_{2}^{*})^{\frac{1}{2}}\sqrt{\alpha_1^{(1)*}}}$, $c_1=\frac{1}{2}\log\frac{(k_1^*-l_1^*)}{(l_1-k_1)}$, $c_2=\frac{1}{2}\log\frac{(k_1-l_1)(k_1^*+l_1)}{(l_1-k_1)(k_1+l_1^*)}$, $e^{i\theta_2^{1-}}=\frac{(l_{1}-l_{2})(l_{1}+l_{2})^{\frac{1}{2}}(l_{1}^*+l_{2})(k_{2}-l_{1})^{\frac{1}{2}}(l_{1}^*-l_{2})^{\frac{1}{2}}(k_{2}+l_{1}^*)^{\frac{1}{2}}\sqrt{\alpha_1^{(2)}}}{(l_{1}^*-l_{2}^*)(l_{1}^*+l_{2}^*)^{\frac{1}{2}}(l_{1}+l_{2}^*)(k_{2}^*-l_{1}^*)^{\frac{1}{2}}(l_{1}-l_{2}^*)^{\frac{1}{2}}(k_{2}^*+l_{1})^{\frac{1}{2}}\sqrt{\alpha_1^{(2)*}}}$, $c_3=\frac{1}{2}\log\frac{(k_1^*-l_1^*)(k_1+l_1^*)}{(k_1^*+l_1)(k_1-l_1)}$, $b_1=(\frac{k_{1R}-l_{1R}}{k_{1R}+l_{1R}})^{\frac{1}{2}}$ $c_4=\frac{1}{2}\log\frac{(k_1^*-l_1^*)(k_1^*+l_1)}{(k_1+l_1^*)(l_1-k_1)}$, and 
 $\phi_{2}^{-}=\frac{1}{2}\ln \frac{-i(k_1-l_1)|k_1-l_2|^2|k_1-k_2|^4|k_1+k_2|^2|\alpha_{1}^{(1)}|^2}{2(k_1-k_1^*)(k_1+k_1^*)^2|k_1+l_2^*|^2(k_1^*+l_1)|k_1-k_2^*|^2|k_1+k_2^*|^4}$. In $e^{i\theta_{1,2}^{1-}}$, superscript ($1-$) represents soliton $1$ ($\mathsf{S}_1$) before collision and  subscripts $(1,2)$ denote the two short-wave components $S^{(1)}$ and $S^{(2)}$, respectively. Note: To get the exact form of ($1,1,1$)-non-degenerate solitons one has to impose the condition: $k_{jI}=l_{jI}$, $j=1,2$, in the constants defined above and the ones appearing in the following asymptotic expressions. \\ 
\underline{Soliton 2}: $\eta_{2R}$, $\xi_{2R}\simeq 0$, $\eta_{1R}$, $\xi_{1R}\rightarrow - \infty$\\
The asymptotic expressions corresponding to soliton 2 are deduced as
\begin{eqnarray}
	&&S^{(1)}\simeq \frac{4k_{2R}\sqrt{k_{2I}}e^{i(\eta_{2I}+\theta_1^{2-})}\cosh(\xi_{2R}+\varphi_1^-)}{\big[a_{21}\cosh(\eta_{2R}+\xi_{2R}+\varphi_1^-+\varphi_2^-+d_1)+a_{21}^{*-1}\cosh(\eta_{2R}-\xi_{2R}+\varphi_2^--\varphi_1^-+d_2)\big]},\nonumber\\
	&&S^{(2)}\simeq \frac{4l_{2R}\sqrt{k_{2I}}e^{i(\xi_{2I}+\theta_2^{2-})}\cosh(\eta_{2R}+\varphi_2^-)}{\big[a_{22}\cosh(\eta_{2R}+\xi_{2R}+\varphi_1^-+\varphi_2^-+d_1)+a_{22}^{*-1}\cosh(\eta_{2R}-\xi_{2R}+\varphi_2^--\varphi_1^-+d_2)\big]},\nonumber\\
	&&L\simeq\frac{4}{f^2}\bigg((k_{2R}^2-l_{2R}^2)+l_{2R}^2\cosh(2\eta_{2R}+2\vphi_2^-+d_3)+k_{2R}^2\cosh(2\xi_{2R}+2\vphi_1^-+d_4)\bigg),\nonumber\\
	&&f=b_2\cosh(\eta_{2R}+\xi_{2R}+\varphi_1^-+\varphi_2^-+d_1)+b_2^{-1}\cosh(\eta_{2R}-\xi_{2R}+\varphi_2^--\varphi_1^-+d_2).\label{12}
\end{eqnarray}
In the above, 
$a_{21}=\frac{(k_{2}^{*}-l_{2}^{*})^{\frac{1}{2}}}{(k_{2}^{*}+l_{2})^{\frac{1}{2}}}$,  $a_{22}=\frac{(k_{2}^{*}-l_{2}^{*})^{\frac{1}{2}}}{(k_{2}+l_{2}^*)^{\frac{1}{2}}}$, 
$e^{i\theta_1^{2-}}=(\frac{\alpha_2^{(1)}}{\alpha_2^{(1)*}})^{\frac{1}{2}}$,
$e^{i\theta_2^{2-}}=(\frac{\alpha_2^{(2)}}{\alpha_2^{(2)*}})^{\frac{1}{2}}$,
 $b_2=\frac{(k_{2R}-l_{2R})^{\frac{1}{2}}}{(k_{2R}+l_{2R})^{\frac{1}{2}}}$,
 $\varphi_1^{-}=\frac{1}{2}\ln\frac{i(k_2-l_2)|\alpha_2^{(2)}|^2}{2(k_2+l_2^*)(l_2-l_2^*)(l_2+l_2^*)^2}$,  $\varphi_2^{-}=\frac{1}{2}\ln\frac{-i(k_2-l_2)|\alpha_2^{(1)}|^2}{2(k_2^*+l_2)(k_2-k_2^*)(k_2+k_2^*)^2}$,
  $d_1=\frac{1}{2}\log\frac{(k_2^*-l_2^*)}{(k_2-l_2)}$, $d_2=\frac{1}{2}\log\frac{(k_2^*+l_2)}{(k_2+l_2^*)}$, $d_3=\frac{1}{2}\log\frac{(k_2^*-l_2^*)(k_2+l_2^*)}{(k_2^*+l_2)(k_2-l_2)}$ and $d_4=\frac{1}{2}\log\frac{(k_2^*-l_2^*)(k_2^*+l_2)}{(k_2+l_2^*)(k_2-l_2)}$.
Here, superscript ($2-$) appears in $e^{i\theta_{1,2}^{2-}}$ refers to soliton 2 ($\mathsf{S}_2$) before collision. \\
\underline{(b) After collision}: $t\rightarrow +\infty$\\
\underline{Soliton 1}: $\eta_{1R}$, $\xi_{1R}\simeq 0$, $\eta_{2R}$, $\xi_{2R}\rightarrow - \infty$\\
The following asymptotic forms are deduced for soliton $\mathsf{S}_1$ after collision from the soliton solution (\ref{A.2a})-(\ref{A.2d}) for $N=2$. The corresponding expressions are
\begin{eqnarray}
	&&S^{(1)}\simeq \frac{4k_{1R}\sqrt{k_{1I}}e^{i(\eta_{1I}+\theta_1^{1+})}\cosh(\xi_{1R}+\phi_1^+)}{\big[{a_{11}}\cosh(\eta_{1R}+\xi_{1R}+\phi_1^++\phi_2^++c_1)+a_{11}^{*-1}\cosh(\eta_{1R}-\xi_{1R}+\phi_2^+-\phi_1^++c_2)\big]},\nonumber\\
	&&S^{(2)}\simeq\frac{4l_{1R}\sqrt{k_{1I}}e^{i(\xi_{1I}+\theta_2^+)}\cosh(\eta_{1R}+\phi_2^+)}{\big[a_{12}\cosh(\eta_{1R}+\xi_{1R}+\phi_1^++\phi_2^++c_1)+a_{12}^{*-1}\cosh(\eta_{1R}-\xi_{1R}+\phi_2^+-\phi_1^++c_2)\big]},\nonumber\\
	&&L\simeq\frac{4}{f^2}\bigg((k_{1R}^2-l_{1R}^2)+l_{1R}^2\cosh(2\eta_{1R}+2\phi_2^++c_3)+k_{1R}^2\cosh(2\xi_{1R}+2\phi_1^++c_4)\bigg),\nonumber\\
	&&f=b_1\cosh(\eta_{1R}+\xi_{1R}+\phi_1^++\phi_2^++c_1)+b_1^{-1}\cosh(\eta_{1R}-\xi_{1R}+\phi_2^+-\phi_1^++c_2).\label{13}
\end{eqnarray}
Here, $\phi_1^+=\frac{1}{2}\ln\frac{i(k_1-l_1)|\alpha_{1}^{(2)}|^2}{2(k_1+l_1^*)(l_1-l_1^*)(l_1+l_1^*)^2}$, $\phi_2^+=\frac{1}{2}\ln\frac{-i(k_1-l_1)|\alpha_{1}^{(1)}|^2}{2(k_1^*+l_1)(k_1-k_1^*)(k_1+k_1^*)^2}$, $e^{i\theta_1^{1+}}=[\alpha_{1}^{(1)}/\alpha_{1}^{(1)^*}]^{1/2}$, and  $e^{i\theta_2^{1+}}=[\alpha_{1}^{(2)}/\alpha_{1}^{(2)^*}]^{1/2}$. In the latter, superscript ($1+$) appears in $e^{i\theta_{1,2}^{1+}}$ represents soliton  $\mathsf{S}_1$ after collision and subscripts $(1,2)$ denote the two SW components $S^{(1)}$ and $S^{(2)}$, respectively. \\ 
\underline{Soliton 2}: $\eta_{2R}$, $\xi_{2R}\simeq 0$, $\eta_{1R}$, $\xi_{1R}\rightarrow + \infty$\\
The asymptotic expressions for soliton $\mathsf{S}_2$ are obtained as
\begin{eqnarray}
	&&S^{(1)}\simeq \frac{4k_{2R}\sqrt{k_{2I}}e^{i(\eta_{2I}+\theta_1^{2+})}\cosh(\xi_{2R}+\varphi_1^+)}{\big[a_{21}\cosh(\eta_{2R}+\xi_{2R}+\varphi_1^++\varphi_2^++d_1)+a_{21}^{*-1}\cosh(\eta_{2R}-\xi_{2R}+\varphi_2^+-\varphi_1^++d_2)\big]},\nonumber\\
	&&S^{(2)}\simeq \frac{4l_{2R}\sqrt{k_{2I}}e^{i(\xi_{2I}+\theta_2^{2+})}\cosh(\eta_{2R}+\varphi_2^+)}{\big[a_{22}\cosh(\eta_{2R}+\xi_{2R}+\varphi_1^++\varphi_2^++d_1)+a_{22}^{*-1}\cosh(\eta_{2R}-\xi_{2R}+\varphi_2^+-\varphi_1^++d_2)\big]},\nonumber\\
	&&L\simeq\frac{4}{f^2}\bigg((k_{2R}^2-l_{2R}^2)+l_{2R}^2\cosh(2\eta_{2R}+2\varphi_1^++d_3)+k_{2R}^2\cosh(2\xi_{2R}+2\varphi_2^++d_4)\bigg),\nonumber\\
	&&f=b_2\cosh(\eta_{2R}+\xi_{2R}+\varphi_1^++\varphi_2^++d_1)+b_2^{-1}\cosh(\eta_{2R}-\xi_{2R}+\varphi_2^+-\varphi_1^++d_2).\label{14}
\end{eqnarray}
Here, $\varphi_1^{+}=\varphi_1^{-}+\frac{1}{2}\ln\frac{|k_1-l_2|^2|l_1-l_2|^4|l_1+l_2|^2}{|k_1+l_2^*|^2|l_1-l_2^*|^2|l_1+l_2^*|^4}$, $e^{i\theta_1^{2+}}=\frac{(k_1-k_2)(k_1+k_2)^{\frac{1}{2}}(k_2-l_1)^{\frac{1}{2}}(k_1-k_2^*)^{\frac{1}{2}}(k_2^*+l_1)^{\frac{1}{2}}(k_1+k_2^*)\alpha_{2}^{(1)\frac{1}{2}}}{(k_1^*-k_2^*)(k_1^*+k_2^*)^{\frac{1}{2}}(k_2^*-l_1^*)^{\frac{1}{2}}(k_1^*-k_2)^{\frac{1}{2}}(k_2+l_1^*)^{\frac{1}{2}}(k_1^*+k_2)\alpha_{2}^{(1)*\frac{1}{2}}}$, $\varphi_2^{+}=\varphi_2^{-}+\frac{1}{2}\ln\frac{|k_1-k_2|^4|k_1+k_2|^2|k_2-l_1|^2}{|k_1-k_2^*|^2|k_1+k_2^*|^4|k_2+l_1^*|^2}$, and 
 $e^{i\theta_2^{2+}}=\frac{(l_1-l_2)(l_1+l_2)^{\frac{1}{2}}(k_1-l_2)^{\frac{1}{2}}(l_1-l_2^*)^{\frac{1}{2}}(k_1+l_2^*)^{\frac{1}{2}}(l_1+l_2^*)\alpha_{2}^{(2)\frac{1}{2}}}{(l_1^*-l_2^*)(l_1^*+l_2^*)^{\frac{1}{2}}(k_1^*-l_2^*)^{\frac{1}{2}}(l_1^*-l_2)^{\frac{1}{2}}(k_1^*+l_2)^{\frac{1}{2}}(l_1^*+l_2)\alpha_{2}^{(2)*\frac{1}{2}}}$. In the latter, superscript ($2+$) arises in $e^{i\theta_{1,2}^{2+}}$ represents soliton  $\mathsf{S}_2$ after collision and subscripts $(1,2)$ denote the two SW components $S^{(1)}$ and $S^{(2)}$, respectively.

The phase constants $\phi_j^-$, $\varphi_j^-$, $\phi_j^+$, $\vphi_j^+$, for $j=1,2$, corresponding to the solitons $\mathsf{S}_1$  and $\mathsf{S}_2$ before and after collision, are related as follows: \begin{subequations}
	\begin{equation}
		\hspace{-2.5cm} \phi_1^+=\phi_1^--\psi_1,~\phi_2^+=\phi_2^--\psi_2,~\vphi_1^+=\vphi_1^-+\Psi_1,~\vphi_2^+=\vphi_2^-+\Psi_2, 
		\label{15a}
	\end{equation}
	where 
	\begin{eqnarray}
		&&\psi_1=\frac{1}{2}\ln\frac{|k_2-l_1|^2|l_1-l_2|^4|l_1+l_2|^2}{|k_2+l_1^*|^2|l_1+l_2^*|^4|l_1-l_2^*|^2},~ \psi_2=\frac{1}{2}\ln\frac{|k_1-k_2|^4|k_1+k_2|^2|k_1-l_2|^2}{|k_1+k_2^*|^4|k_1-k_2^*|^2|k_1+l_2^*|^2},\nonumber\\ 
		&&\Psi_1=\frac{1}{2}\ln\frac{|k_1-l_2|^2|l_1-l_2|^4|l_1+l_2|^2}{|k_1+l_2^*|^2|l_1+l_2^*|^4|l_1-l_2^*|^2},~\Psi_2=\frac{1}{2}\ln\frac{|k_2-l_1|^2|k_1-k_2|^4|k_1+k_2|^2}{|k_2+l_1^*|^2|k_1+k_2^*|^4|k_1-k_2^*|^2}.\label{15b}	
\end{eqnarray} \end{subequations}
From the above, it is evident that the phases of each soliton are altered during the collision dynamics. The total phase shift experienced by soliton $\mathsf{S}_1$  in both the SW components is given by:
\begin{subequations}
	\begin{eqnarray}
		\hspace{-1.5cm}\Delta \Phi_1&=&\log \frac{|k_2-l_1||l_1-l_2|^2|l_1+l_2||k_1-l_2||k_1-k_2|^2|k_1+k_2|}{|k_2+l_1^*||l_1+l_2^*|^2|l_1-l_2^*||k_1+l_2^*||k_1+k_2^*|^2|k_1-k_2^*|}.
	\end{eqnarray}
The total phase shift experienced by soliton $\mathsf{S}_2$ in the SW components can be expressed as:
	\begin{eqnarray}
		\hspace{-1.5cm}\Delta \Phi_2&=&-\log \frac{|k_2-l_1||l_1-l_2|^2|l_1+l_2||k_1-l_2||k_1-k_2|^2|k_1+k_2|}{|k_2+l_1^*||l_1+l_2^*|^2|l_1-l_2^*||k_1+l_2^*||k_1+k_2^*|^2|k_1-k_2^*|}=-\Delta \Phi_1.~~~~~~~
	\end{eqnarray}
\end{subequations}
The subscript $1$ and $2$ appearing in $\Delta \Phi$ denote the soliton number. The total phase shifts calculated for the SW components are the same as those for the LW component.
\begin{figure}
	\centering\includegraphics[width=1.0\linewidth]{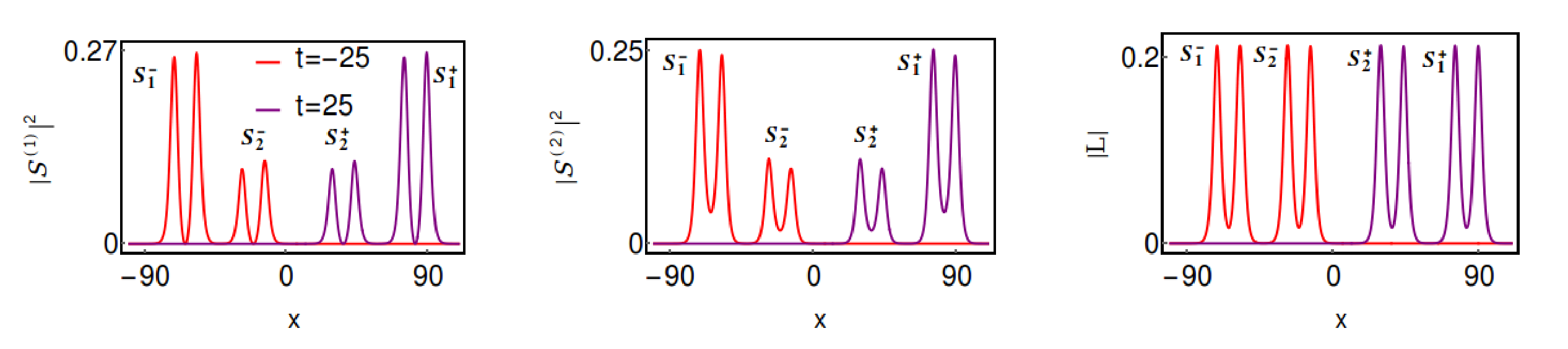} 
	\caption{Shape preserving collision between two asymmetric double-hump solitons for the parametric values $\sigma_1=\sigma_2=-1$, $k_1=0.32+0.5i$, $l_1=0.333+0.5i$, $k_2=0.32+1.2i$, $l_2=0.333+1.2i$, $\alpha_1^{(1)}=0.6+0.5i$, $\alpha_1^{(2)}=0.61+0.5i$, $\alpha_2^{(1)}=0.62+0.55i$, and $\alpha_2^{(2)}=0.61+0.5i$.}
	\label{fig3}
\end{figure}
\subsubsection{Elastic collision property of $(1,1,1)$ non-degenerate solitons}
\begin{figure}
	\centering\includegraphics[width=1.0\linewidth]{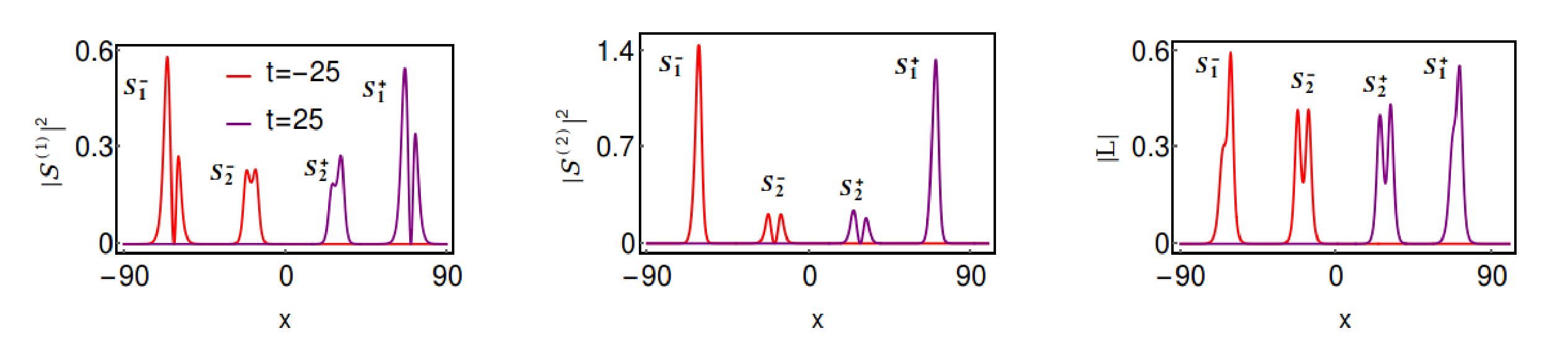} \includegraphics[width=1.0\linewidth]{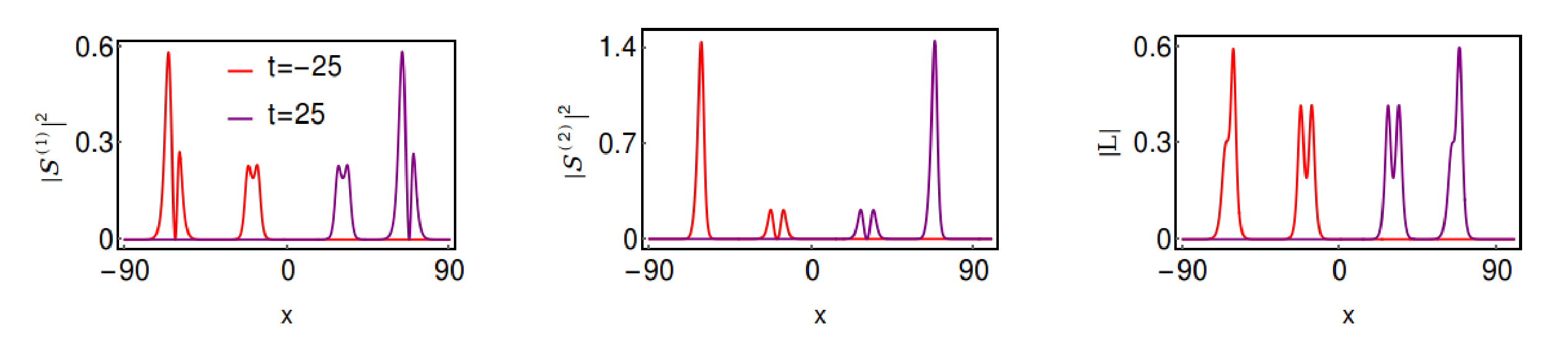}
	\caption{Top-Panels: Shape changing collision between two non-degenerate solitons for the parametric values $\sigma_1=\sigma_2=-1$, $k_1=0.515+0.5i$, $l_1=0.412+0.5i$, $k_2=0.35+1.2i$, $l_2=0.6+1.2i$, $\alpha_1^{(1)}=0.45+0.5i$, $\alpha_1^{(2)}=0.65+0.65i$, $\alpha_2^{(1)}=1.2+1.5i$, and $\alpha_2^{(2)}=0.7+0.7i$. Bottom-Panels: We demonstrate that the shape-changing collision remains an elastic collision by restoring the solitons to their shape-preserving form. This is achieved by applying the time shifts given in Eq. (\ref{16}). }
	\label{fig4}
\end{figure}

The asymptotic analysis indicates that while the amplitudes of non-degenerate solitons remain constant during collision process, their phase terms vary. As a result, $(1,1,1)$ non-degenerate solitons are expected to undergo the standard elastic collision in the system (\ref{eq.1}), for negative nonlinearity $\sigma_l<0$, $l=1,2$,  with a finite phase shift. Nevertheless, as shown in Fig. \ref{fig4}, they also experience shape-changing collision apart from a shape-preserving collision. This interesting behavior of $(1,1,1)$ non-degenerate solitons in the present system (\ref{eq.1}) can be understood only by analysing the variations in the phase shifts. From Eqs. (\ref{15a}) and (\ref{15b}), it can be seen that the phase shifts before the collision, denoted by $\phi_j^-$ and $\varphi_j^-$, change after the collision to $\phi_j^+=\phi_j^--\psi_j$ and $\varphi_j^+=\varphi_j^-+\Psi_j$, $j=1,2$. Therefore, these additional phase shifts, $\psi_j$ and $\Psi_j$, $j=1,2$, are responsible for the shape-changing behavior of $(1,1,1)$ non-degenerate solitons observed in Fig. \ref{fig4}. On the other hand, if these additional phase shifts vanish, the solitons can preserve their structures and undergo collision  with a zero phase shift. A condition for realizing such a shape-preserving collision with zero phase shift can be deduced by setting $\psi_j=0$ and  $\Psi_j=0$, for $j=1,2$. This requires the arguments of the logarithmic functions in both $\psi_j$'s and  $\Psi_j$'s to be equal to one. By doing so, the following zero phase shift condition is derived, which reads as:
\begin{equation}
	\frac{|k_2-l_1|^2}{|k_2+l_1^*|^2}-\frac{|k_1-l_2|^2}{|k_1+l_2^*|^2}=0.\label{15}
\end{equation} 
The above criterion implies that whenever the wave numbers associated with the $(1,1,1)$ non-degenerate solitons satisfy this condition, they undergo shape-preserving collisions with zero phase shift. Under this situation, the $(1,1,1)$ non-degenerate solitons can pass through each other while completely maintaining their structures. We demonstrate below the distinction between the shape-preserving and shape-changing behaviors of $(1,1,1)$-non-degenerate solitons, and explain how both cases correspond to elastic collision.      

To start with, we consider the shape preserving nature of $(1,1,1)$-non-degenerate solitons. It is brought out in Fig. \ref{fig3} by setting the paramater values, $k_1=0.32+0.5i$, $l_1=0.333+0.5i$, $k_2=0.32+1.2i$, $l_2=0.333+1.2i$, $\alpha_1^{(1)}=0.6+0.5i$, $\alpha_1^{(2)}=0.61+0.5i$, $\alpha_2^{(1)}=0.62+0.55i$, and $\alpha_2^{(2)}=0.61+0.5i$. Figure \ref{fig3} clearly illustrates that the two double-hump solitons maintain their shapes in all the components after the collision, thereby confirming the elastic nature of the interaction. We have also verified that the aforementioned numerical values of the wave numbers do obey the constraint Eq. (\ref{15}). Therefore, the collision which is demonstrated in Fig. \ref{fig3} is nothing but shape-preserving collision.
However, in general, the phase terms vary significantly, which in turn affects the structures of the $(1,1,1)$ non-degenerate solitons. As a result, depending on the values of the additional phase shifts, the solitons may undergo either slight or substantial structural changes in their profiles. A typical example of a novel shape-changing collision is illustrated in Fig. \ref{fig4} for the following values: $k_1=0.515+0.5i$, $l_1=0.412+0.5i$, $k_2=0.35+1.2i$, $l_2=0.6+1.2i$, $\alpha_1^{(1)}=0.45+0.5i$, $\alpha_1^{(2)}=0.65+0.65i$, $\alpha_2^{(1)}=1.2+1.5i$, and $\alpha_2^{(2)}=0.7+0.7i$. As it is evident from Fig. \ref{fig4}, the shape-changing behavior is observed in both the SW components and the LW component. This is in total contrast to the behavior of degenerate solitons, where shape changes occur only in the SW components, along with energy redistribution \cite{kanna1}. However, to visualize that the shape-changing collision corresponds to an elastic collision, we consider a pair of time shifts in the asymptotic expressions (Eqs. (\ref{13}) and (\ref{14})) of the $(1,1,1)$-non-degenerate solitons. That is, the shape-changing effect can be undone by applying the time shifts
\begin{eqnarray}
	\big(t'=t+\frac{\psi_1}{2l_{1R}k_{1I}}, ~t'=t+\frac{\psi_2}{2k_{1R}k_{1I}}\big)~~ \text{and}~~ \big(t'=t-\frac{\Psi_1}{2l_{2R}k_{2I}}, ~t'=t-\frac{\Psi_2}{2k_{2R}k_{2I}}\big), \label{16}
\end{eqnarray}
in the expressions (\ref{13}) and  (\ref{14}), respectively. By incorporating these time shifts into the respective asymptotic expressions, we find that the non-degenerate solitons retain their shapes asymptotically, exhibiting zero phase shift, as graphically demonstrated in the bottom panels of Fig. \ref{fig4}. This confirms the elastic nature of the collision. 

We note that the collision behavior of the  $(1,1,1)$ non-degenerate solitons in the system (\ref{eq.1}) with positive nonlinearity coefficients $\sigma_l>0$ is similar to the case discussed above for the negative nonlinearity ($\sigma_l<0$) case, except that the collision occurs in the $-x$-direction. In fact, this scenario has already been analyzed in our earlier work \cite{stalin-lsri}. For a mixed choice of nonlinearity coefficients, the LSRI system (\ref{eq.1}) admits singular higher-order soliton solutions, similar to the case of the one-soliton solution. As a result, analyzing such higher-order solutions is physically meaningless, and therefore, this possibility is excluded from our analysis.
\subsubsection{Elastic collision of $(1,1,2)$ non-degenerate solitons}
\begin{figure}
	\centering\includegraphics[width=1.0\linewidth]{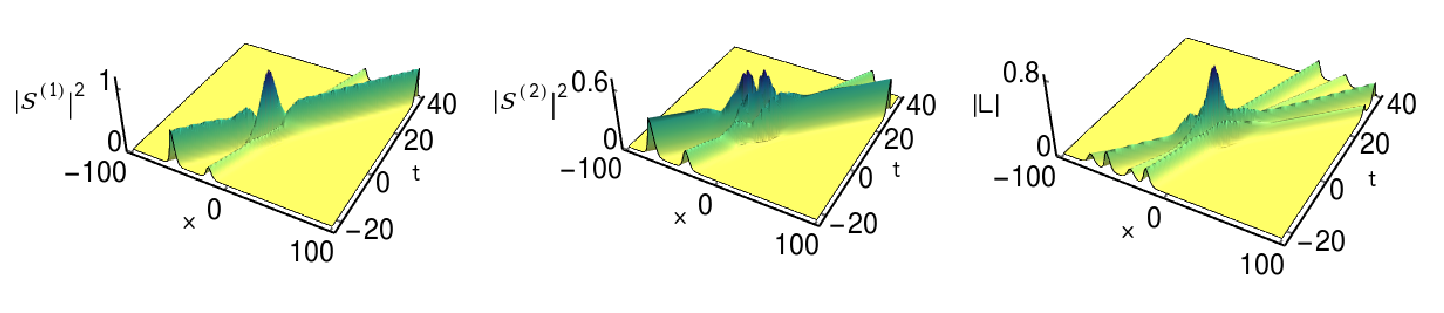} 
	\caption{Elastic collision between two $(1,1,2)$ non-degenerate solitons: $\sigma_1=\sigma_2=-1$, $k_1=0.415+0.5i$, $l_1=0.4+0.5i$, $k_2=0.3+1.2i$, $l_2=0.315+1.2i$, $\alpha_1^{(1)}=0.6+0.5i$, $\alpha_1^{(2)}=0.61+0.5i$, $\alpha_2^{(1)}=0.62+0.55i$, and $\alpha_2^{(2)}=0.61+0.5i$.}
	\label{fig5}
\end{figure}

We next investigate the interaction dynamics of $(1,1,2)$ non-degenerate solitons by by setting $N=2$ in the soliton solution (\ref{A.2a})-(\ref{A.2d}), with the velocity condition  $k_{jI}\neq l_{jI}$, $j=1,2$. The resulting solution under this choice is referred as the ($2,2,4$)-soliton solution, which consists of two distinct $(1,1,2)$ non-degenerate solitons. In this case, two distinct sets of single-humped solitons appear in both the SW components $S^{(1)}$ and $S^{(2)}$, while in the LW component, $L$, these solitons reemerge, increasing the total number of solitons to four. Figure \ref{fig5} presents a typical example of the collision process involving $(1,1,2)$ non-degenerate solitons in the system (\ref{eq.1}) with $\sigma_l<0$. As we can see from Fig. \ref{fig5}, it is clear that the structures of $(1,1,2)$ non-degenerate solitons remain constant throughout the collision process, thereby confirming the elastic nature of the interaction. This type of elastic collision can also be observed in the system (\ref{eq.1}) with positive nonlinearity coefficients $\sigma_l>0$, as well as in the mixed nonlinear LSRI system (\ref{eq.1}). However, in the mixed LSRI system, the elastic collision occurs between oppositely moving $(1,1,2)$ non-degenerate solitons.
\subsubsection{Collision between $(1,1,1)$ and $(1,1,2)$ non-degenerate solitons}
\begin{figure}
	\centering\includegraphics[width=1.0\linewidth]{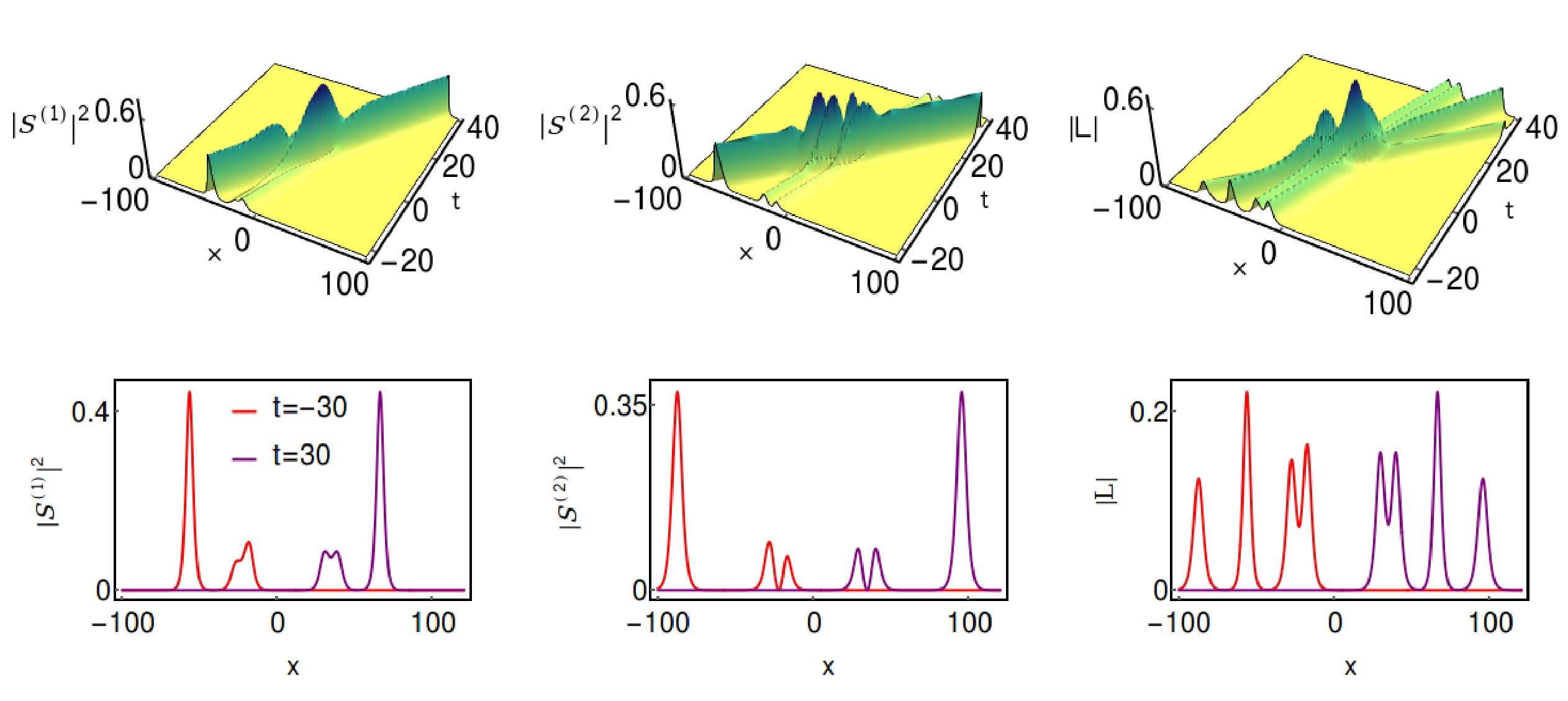} 
	\caption{Shape-changing collision of $(1,1,1)$ non-degenerate soliton induced by $(1,1,2)$ non-degenerate soliton. }
	\label{fig6}
\end{figure}

To study the collision scenario between $(1,1,1)$ and $(1,1,2)$ non-degenerate solitons, we consider the condition, $k_{1I}= l_{1I}$ and $k_{2I}\neq l_{2I}$, in the soliton solution (\ref{A.2a})-(\ref{A.2d}), for $N=2$. The resultant solution is displayed in Fig. \ref{fig6} for the parameter values,  $\sigma_1=\sigma_2=-1$, $k_1=0.315+0.5i$, $l_1=0.25+0.5i$, $k_2=0.333+i$, $l_2=0.25+1.5i$, $\alpha_1^{(1)}=0.45+0.5i$, $\alpha_1^{(2)}=0.5+0.5i$, $\alpha_2^{(1)}=0.45+0.5i$, and $\alpha_2^{(2)}=0.5+0.5i$.  Figure \ref{fig6} shows that the $(1,1,2)$ non-degenerate soliton induces shape changing property of  $(1,1,1)$ non-degenerate soliton. That is, asymmetric double-humped soliton becomes symmetric in all the components when it collides with $(1,1,2)$ non-degenerate soliton. We have investigated this possibility through asymptotic analysis, but the corresponding results are omitted here for brevity.     


\section{\bf Collision between degenerate and non-degenerate solitons}
To investigate the shape-changing behavior of a non-degenerate soliton induced by its collision with a degenerate soliton, we consider a special form of the two-soliton solution that includes both degenerate and non-degenerate solitons. Such a two-soliton solution can be obtained by applying a partial non-degenerate (or partial degenerate) limit: $k_1\neq l_1$ and $k_2=l_2$, in Eqs. (\ref{A.2a})-(\ref{A.2d}) with $N=2$. The resulting solution is referred to as the partial non-degenerate two-soliton solution. Consequently, it becomes important to investigate the collision dynamics between degenerate and non-degenerate solitons. As we know, the non-degenerate fundamental soliton exists in two distinct forms, leading to two possible types of collision dynamics: (i) collision between a degenerate soliton and a ($1,1,1$)-non-degenerate soliton, and (ii) collision between a degenerate soliton and a ($1,1,2$)-non-degenerate soliton. However, we intend to analyse the former only for brevity. 

 To explore the degenerate bright soliton collision induced shape changing behaviour of the ($1,1,1$) non-degenerate soliton, we intend to analyse the partial non-degenerate two-soliton solution in the limits $t\rightarrow \pm \infty$. In these limits, the asymptotic forms associated with the non-degenerate and degenerate solitons can be deduced by substituting the respective asymptotic behavior of their corresponding wave variables $\eta_{1R}=k_{1R}(t-2k_{I1}z)$, $\xi_{1R}=l_{1R}(t-2k_{1I}z)$, and $\eta_{2R}=k_{2R}(t-2k_{2I}z)$ in the obtained partially non-degenerate two-soliton solution. In the latter,  the variables $\eta_{1R}$ and $\xi_{1R}$ represent the non-degenerate soliton and  $\eta_{2R}$ denotes the degenerate soliton. To deduce their asymptotic expressions, we consider the parametric choices, $k_{jR},l_{1R}>0$,  $k_{1I}=l_{1I}<k_{2I}$.  For this choice, the wave variables behave asymptotically as follows: (i) ($1,1,1$)-non-degenerate soliton $\mathsf{S}_1$: $\eta_{1R},\xi_{1R}\simeq0$, $\eta_{2R}\rightarrow \pm\infty$ as $t\rightarrow \mp\infty$ (ii) degenerate soliton $\mathsf{S}_2$: $\eta_{2R}\simeq 0$, $\eta_{1R}, \xi_{1R} \rightarrow\pm \infty$ as $t\rightarrow\pm \infty$.We have deduced the following expressions to describe the structural changes that occur in non-degenerate and degenerate solitons during the collision process.\\
 \underline{(a) Before collision}: $t\rightarrow -\infty$\\
  \underline{Non-degenerate soliton}: The asymptotic forms of the nondegenerate soliton $\mathsf{S}_1$, which is present in both the short-wave components as well as in the long-wave component, before collision are obtained as
 \begin{subequations}
 	\bea
 	&&S^{(1)}\simeq\frac{1}{D_1}\big(e^{i\eta_{1I}}d_{11}\cosh(\xi_{1R}+\Phi_1^-)+e^{i\xi_{1I}}d_{12}\cosh(\eta_{1R}+\Phi_2^-)\big),\\
 	&&S^{(2)}\simeq\frac{1}{D_1}\big(e^{i\eta_{1I}}d_{21}\cosh(\xi_{1R}+\Phi_6^-)+e^{i\xi_{1I}}d_{22}\cosh(\eta_{1R}+\Phi_7^-)\big),\\
 	&&L\simeq\frac{1}{D_1^2}\bigg(4k_{1R}^2\nu_{11}\cosh(2\xi_{1R}+\Phi_8^-)+4l_{1R}^2\nu_{11}\cosh(2\eta_{1R}+\Phi_{9}^-)+2(k_{1R}-l_{1R})^2\nu_{12}\cosh\Phi_{5}^-\nonumber\\
 	&&\hspace{0.5cm}\times[\cos\theta\cosh(\eta_{1R}-\xi_{1R}+\Phi_{4}^-)+i\sin\theta\sinh(\eta_{1R}-\xi_{1R}+\Phi_{4}^-)]+2(k_{1R}-l_{1R})^2\nu_{13}\nonumber\\
 	&&\hspace{0.5cm}\times \cosh(\eta_{1R}+\xi_{1R}+\Phi_8^-)e^{\frac{\eta_{1R}+\xi_{1R}}{2}}[\cosh\Phi_5^-\cos\theta+i\sinh\Phi_5^-\sin\theta]+\nu_{14}\bigg),\\
 	&&D_1=d_{13}\cosh(\eta_{1R}+\xi_{1R}+\Phi_3^-)+d_{14}\cosh(\eta_{2R}-\xi_{2R}+\Phi_4^-)\nonumber\\
 	&&\hspace{0.5cm}+d_{15}[\cosh\Phi_5^-\cos\theta+i\sinh\Phi_5^-\sin\theta],
 \end{eqnarray}
\end{subequations}
where $\theta=(k_{1R}^2-l_{1R}^2)t$. The constants such as $d_{ij}$'s, $\nu_{11}$, $\nu_{12}$, $\nu_{13}$, $\nu_{14}$, $\Phi_j^-$, $i=1,2$, $j=1,2,...,9$, appearing above are defined in Appendix B.\\
 \underline{Degenerate soliton}: The asymptotic form of the degenerate soliton, deduced from the partially non-degenerate soliton solution (Eqs. (\ref{A.2a})-(\ref{A.2d}) with $N=2$, $k_l\neq l_1$ and $k_2=l_2$), is given by
 \begin{eqnarray}
 	\hspace{-0.5cm}S^{(l)}\simeq\begin{pmatrix}
 		A_1^{-}\\ 
 		A_2^{-}
 	\end{pmatrix} 2k_{2R}\sqrt{k_{2I}}e^{i(\eta_{2I}-\frac{\pi}{2})}\sech(\eta_{2R}+\phi^-), ~L\simeq 2k_{2R}^2\sech^2(\eta_{2R}+\phi^-),~l=1,2.~~
 	\label{5.1}
 \end{eqnarray}
 where $A_l^{-}=\al_2^{(l)}/(\sigma_1|\al_2^{(1)}|^2+\sigma_2|\al_2^{(2)}|^2)^{1/2}$, $l=1,2$, $\phi^-=\frac{1}{2}\ln\frac{(\sigma_1|\al_2^{(1)}|^2+\sigma_2|\al_2^{(2)}|^2)}{2i(k_2+k_2^*)^2(k_2-k_2^*)}$. Here, the subscript $l$ in $A_l^{-}$  denotes the SW components and superscript ($-$) represents before collision. \\
\underline{(b) After collision}: $t\rightarrow +\infty$\\
\underline{Non-degenerate soliton}: In this limit, the asymptotic forms corresponding to the non-degenerate soliton $\mathsf{S}_1$  after collision are deduced as
\begin{subequations}
\begin{eqnarray}
	&&S^{(1)}\simeq \frac{4k_{1R}\sqrt{k_{1I}}e^{i(\eta_{2I}+\theta_1+\frac{\pi}{2})}\cosh(\xi_{1R}+\Phi_1^+)}{\big[a_{11}\cosh(\eta_{1R}+\xi_{1R}+\Phi_1^++\Phi_2^+)+\frac{1}{a_{11}^*}\cosh(\eta_{1R}-\xi_{1R}+\Phi_2^+-\Phi_1^+)\big]},\\
	&&S^{(2)}\simeq \frac{4l_{1R}\sqrt{k_{1I}}e^{i(\eta_{1I}+\theta_2+\frac{\pi}{2})}\cosh(\eta_{1R}+\Phi_2^+)}{\big[a_{12}\cosh(\eta_{1R}+\xi_{1R}+\Phi_1^++\Phi_2^+)+\frac{1}{a_{12}^*}\cosh(\eta_{1R}-\xi_{1R}+\Phi_2^+-\Phi_1^+)\big]},\\
	&&L\simeq \frac{4}{D_2^2}\bigg(k_{1R}^2\cosh(2\xi_{1R}+2\Phi_1^++\delta)+(k_{1R}^2-l_{R}^2)+l_{1R}^2\cosh(2\eta_{1R}+2\Phi_2^++\delta)\bigg),\\
	&&D_2=\big(b_1\cosh(\eta_{1R}+\xi_{1R}+\Phi_1^++\Phi_2^+)+b_1^{-1}\cosh(\eta_{1R}-\xi_{1R}+\Phi_2^+-\Phi_1^+)\big),\nonumber
\end{eqnarray}
\end{subequations}
where $e^{i\theta_j}=[\alpha_1^{(j)}/\alpha_1^{(j)*}]^{1/2}$, $j=1,2$, $\delta=\frac{1}{2}\ln\frac{|k_1-l_1|^2}{|k_1+l_1|^2}$, $\Phi_1^+=\frac{1}{2}\ln\frac{-i\sigma_2(k_1-l_1)|\alpha_1^{(2)}|^2}{2(k_1+l_1^*)(l_1-l_1^*)(l_1+l_1^*)^2}$, $\Phi_2^+=\frac{1}{2}\ln\frac{i\sigma_1(k_1-l_1)|\alpha_1^{(1)}|^2}{2(k_1^*+l_1)(k_1-k_1^*)(k_1+k_1^*)^2}$, and other constants such as $a_{11}$, $a_{12}$ and $b_1$ are defined below Eq. (\ref{11}). \\
\underline{Degenerate soliton}: The asymptotic expressions for the degenerate soliton $\mathsf{S}_2$ after collision are obtained as
\begin{subequations}
\begin{eqnarray}
	\hspace{-0.5cm}S^{(1,2)}\simeq\begin{pmatrix}
		A_1^{+}\\
		A_2^{+}
	\end{pmatrix}2k_{2R}\sqrt{k_{2I}}e^{i(\eta_{2I}+\frac{\pi}{2})}\sech(\eta_{2R}+\phi^+),~
	L\simeq 2k_{2R}^2\sech^2(\eta_{2R}+\phi^+).~~\label{5.4}
\end{eqnarray}
\end{subequations}
In the above,  $A_1^{+}=\al_2^{(1)}/(\sigma_1|\al_2^{(1)}|^2+\sigma_2\chi|\al_2^{(2)}|^2)^{1/2}$, $A_2^{+}=\al_2^{(1)}/(\sigma_1|\al_2^{(1)}|^2\chi^{-1}+\sigma_2|\al_2^{(2)}|^2)^{1/2}$,\\ $\chi=(|k_2-l_1|^2|k_1+k_2^*|^2|k_2+l_1|^2|k_1-k_2^*|^2)/(|k_1-k_2|^2|k_2+l_1^*|^2|k_1+k_2|^2|k_2-l_1^*|^2)$, $\phi^+=\frac{1}{2}\ln\frac{|k_1-k_2|^2|k_2-l_1|^2\Lam_3}{2i(k_2-k_2^*)(k_2+k_2*)^2|k_1-k_2^*|^2|k_1+k_2^*|^4|k_2-l_1^*|^2|k_2+l_1^*|^4}$, where $\Lambda_3$ is defined in Appendix B. Here, $l$ in $A_l^{+}$, $l=1,2$, refers to the SW components  and the superscript ($+$) denotes after collision.

\begin{figure}
	\centering\includegraphics[width=1.0\linewidth]{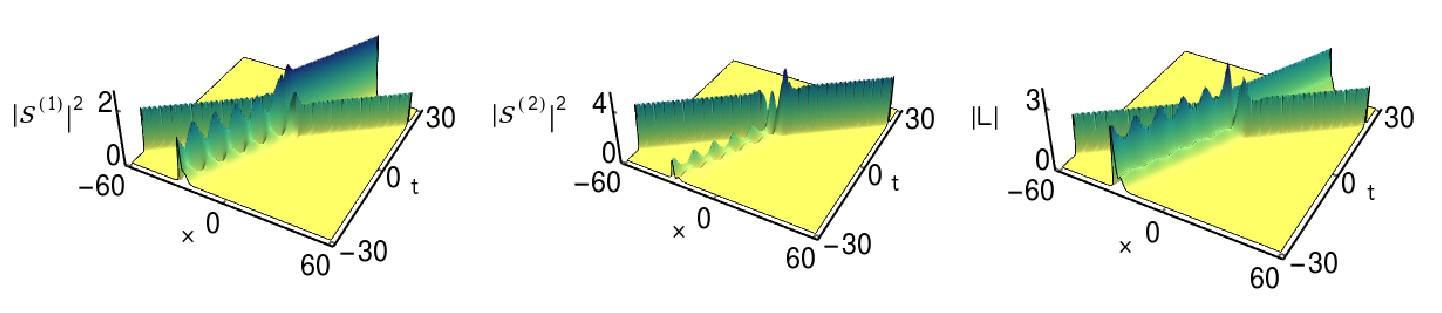} \includegraphics[width=1.0\linewidth]{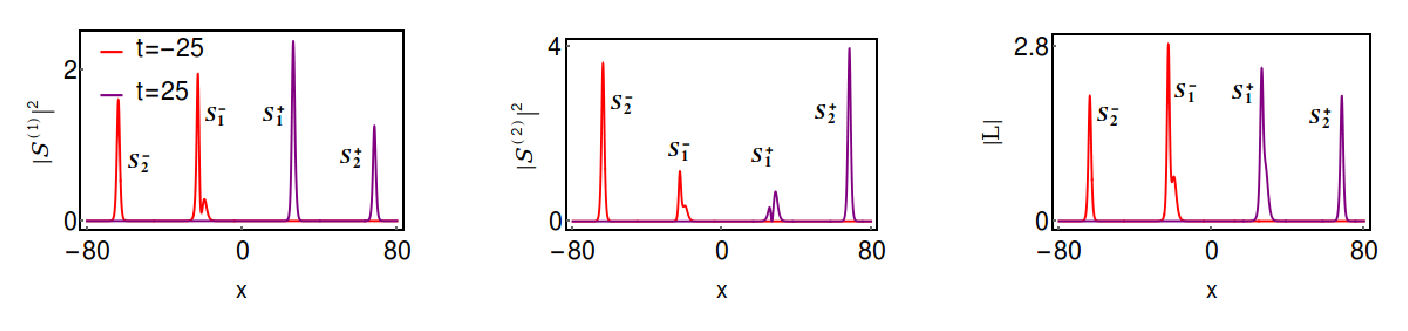}
	\caption{Energy sharing collision between a degenerate soliton and a non-degenerate soliton: $\sigma_1=\sigma_2=-1$, $k_1=1.2+0.5i$, $l_1=0.6+0.5i$, $k_2=l_2=1+1.3i$, $\alpha_1^{(1)}=0.7$, $\alpha_1^{(2)}=0.6$, $\alpha_2^{(1)}=0.8$, and $\alpha_2^{(2)}=1.2$.}
	\label{fig7}
\end{figure}

\begin{figure}
	\centering\includegraphics[width=1.0\linewidth]{ 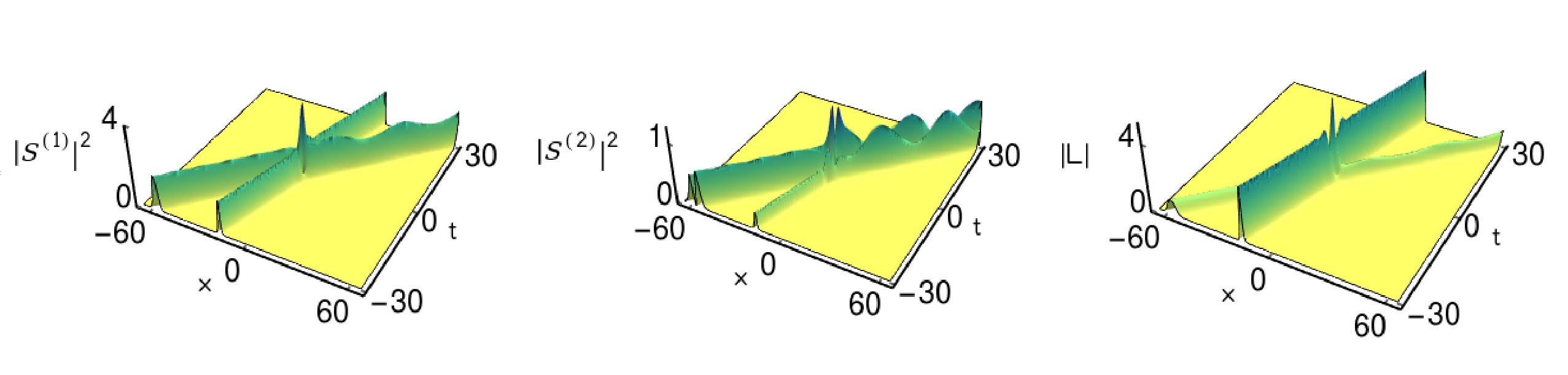} 
	\caption{The collision with a degenerate soliton induces a breathing structure in the non-degenerate soliton: $\sigma_1=\sigma_2=-1$,  $k_1=0.8+i$, $l_1=0.45+i$, $k_2=l_2=1.3+0.3i$, $\alpha_1^{(1)}=0.45$, $\alpha_1^{(2)}=0.45+0.55i$, $\alpha_2^{(1)}=1+i$, and $\alpha_2^{(2)}=0.5$.}
	\label{fig7a}
\end{figure}

From the above analysis, we find that the degenerate soliton $\mathsf{S}_2$ strongly influences the structure of ($1,1,1$)-non-degenerate soliton as it is demonstrated in Figs. \ref{fig7} and \ref{fig7a}. As a result, the geometrical structure of non-degenerate soliton $\mathsf{S}_1$ gets varied to a different profile structure.  For instance, we display such a typical energy sharing collision between a single-humped degenerate soliton and an asymmetric breathing ($1,1,1$)-non-degenerate soliton in Fig. \ref{fig7}. As we can see from this figure, the degenerate soliton $\mathsf{S}_2$ undergoes the standard energy sharing collision among the two SW components while colliding with a non-degenerate soliton. That is, the intensity of soliton $\mathsf{S}_2$ is suppressed in the SW component $S^{(1)}$ whereas it is enhanced in the other SW component $S^{(2)}$. To preserve the energy conservation, the intensity of soliton $\mathsf{S}_1$ is enhanced apprecibly accompanied by a profile transition from a breathing asymmetric shape to a single-humped form with a standard phase shift. In the LW component, the degenerate soliton retains its structural profile, as observed in the purely degenerate case \cite{kanna1}, whereas the ($1,1,1$)-non-degenerate soliton $\mathsf{S}_1$ exhibits a distinct shape-changing behavior. This novel shape changing collision has not been reported before in the literature for the system (\ref{eq.1}) with a negative nonlinearity coefficient. During this interesting enegy sharing collision, the  polarization constants corresponding to the degenerate soliton of SW components before collision $A_l^{-}=\al_2^{(l)}/(\sigma_1|\al_2^{(1)}|^2+\sigma_2|\al_2^{(2)}|^2)^{1/2}$, $l=1,2$,  change into $A_1^{+}=\al_2^{(1)}/(\sigma_1|\al_2^{(1)}|^2+\sigma_2\chi|\al_2^{(2)}|^2)^{1/2}$, $A_2^{+}=\al_2^{(2)}/(\sigma_1|\al_2^{(1)}|^2\chi^{-1}+\sigma_2|\al_2^{(2)}|^2)^{1/2}$, where  $\chi=(|k_2-l_1|^2|k_1+k_2^*|^2|k_2+l_1|^2|k_1-k_2^*|^2)/(|k_1-k_2|^2|k_2+l_1^*|^2|k_1+k_2|^2|k_2-l_1^*|^2)$. 
 
\begin{figure}
	\centering\includegraphics[width=1.0\linewidth]{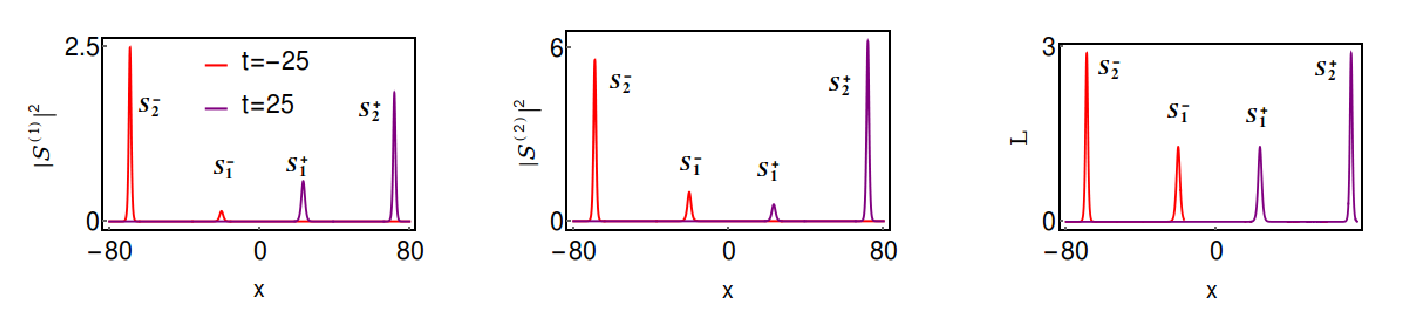} 
	\caption{Energy sharing collision between two degenerate solitons: $\sigma_1=\sigma_2=-1$, $k_1=l_1=0.8+0.45i$, $k_2=l_2=1.2+1.4i$, $\alpha_1^{(1)}=\alpha_1^{(2)}=0.8$, $\alpha_2^{(1)}=1$, and $\alpha_2^{(2)}=1.5$.}
	\label{fig8}
\end{figure}

To characterize this energy sharing collision, one can calculate the transition amplitudes corresponding to the degenerate soliton $\mathsf{S}_2$. The explicit forms of the transition amplitudes are calculated as \begin{eqnarray}
	T_1=\frac{(\sigma_1|\al_2^{(1)}|^2+\sigma_2|\al_2^{(2)}|^2)^{1/2}}{(\sigma_1|\al_2^{(1)}|^2+\sigma_2\chi|\al_2^{(2)}|^2)^{1/2}},~T_2=\frac{(\sigma_1|\al_2^{(1)}|^2+\sigma_2|\al_2^{(2)}|^2)^{1/2}}{(\sigma_1|\al_2^{(1)}|^2\chi^{-1}+\sigma_2|\al_2^{(2)}|^2)^{1/2}}, \label{24}
\end{eqnarray}
where $\chi=(|k_2-l_1|^2|k_1+k_2^*|^2|k_2+l_1|^2|k_1-k_2^*|^2)/(|k_1-k_2|^2|k_2+l_1^*|^2|k_1+k_2|^2|k_2-l_1^*|^2)$.  In general, the value of $\chi$ is not equal to one. As a result, the transition amplitudes $T_1$ and $T_2$ are no longer unimodular. This leads to the occurrence of shape-changing collisions between the degenerate and ($1,1,1$) non-degenerate interacting solitons. To bring out the standard elastic collision one has to choose the wave numbers appropriately to set  $\chi=1$. In this situation, the quantities $T_1$ and $T_2$ become unimodular. We note that the position shifts that occured during the collision process can be calculated from the asymptotic expressions of degenerate and non-degenerate solitons. Note that one can also bring out the breathing nature of ($1,1,1$)-non-degenerate soliton by allowing it to collide with a degenerate soliton. We illustrate this possibility in Fig. \ref{fig7a} for the parameter values, $k_1=0.8+i$, $l_1=0.45+i$, $k_2=l_2=1.3+0.3i$, $\alpha_1^{(1)}=0.45$, $\alpha_1^{(2)}=0.45+0.55i$, $\alpha_2^{(1)}=1+i$, and $\alpha_2^{(2)}=0.5$. In this case also, the degenerate soliton exhibits its characteristic energy-sharing collision behavior, whereas the initially stable, non-breathing profile of the $(1,1,1)$-non-degenerate soliton evolves into a novel breathing structure due to the collision.    

For completeness, we also demonstrate the energy sharing collision of the degenerate solitons of the LSRI system (\ref{eq.1}).  Such a typical collision scenario is displayed in Fig. \ref{fig8} as an example. From this figure, one can easily observe that the energy of the soliton $\mathsf{S}_2$ is suppressed  in the $S^{(1)}$ component and it gets enhanced in the $S^{(2)}$ component. In order to preserve the conservation of energy in both the SW components, the energy of the soliton $\mathsf{S}_2$ is enhanced in the  $S^{(1)}$ component and it gets suppressed in the  $S^{(2)}$ component. However, the degenerate solitons always  undergo elastic collision in the long-wave component. The total energies of each of the degenerate solitons are conserved among the components. The elastic collision is brought out in all the components by fixing the parameters as $\frac{\alpha_1^{(1)}}{\alpha_2^{(1)}}=\frac{\alpha_1^{(2)}}{\alpha_2^{(2)}}$ \cite{kanna1}.


\section{Conclusions}\label{sec-conclu} 
We have demonstrated the emergence of non-degenerate bright solitons in a completely integrable two-component long-wave--short-wave resonance interaction model with a general form of nonlinearity parameters and summarized their several interesting features.  For this purpose, we have adopted the classical Hirota's bilinear method and obtained a fully non-degenerate $N$-soliton solution for this two-component LSRI model in order to analyse the nature of non-degenerate vector solitons in detail. The fundamental non-degenerate soliton is classified as ($1,1,1$)- and ($1,1,2$)-non-degenerate solitons with respect to the velocity conditions. We then showed that depending on the choice of wave parameters, the basic ($1,1,1$)-non-degenerate soliton can exhibit five distinct profile structures, including a novel double-hump, a special flat-top, and a conventional single-hump profile. On the other hand, the ($1,1,2$)-non-degenerate soliton exhibits a two-soliton-like collision behavior in the LW component. Further, we have carried out an appropriate asymptotic analysis to verify the collision nature of ($1,1,1$)-non-degenerate solitons. The asymptotic behavior of these non-degenerate solitons revealed that they can exhibit both shape-preserving and shape-changing collisions. However, our study confirmed that the shape changing collision can be visualized as an elastic collision by taking appropriate time shifts. On the other hand, ($1,1,2$)-non-degenerate solitons straightaway undergo elastic collision with a standard phase shift. Further, we have investigated the formation or suppression of breathing phenomena during collision between a degenerate soliton and a non-degenerate soliton by considering partially non-degenerate multi-soliton solutions. Finally, for completeness, we have also pointed out the energy sharing collision scenario among the two degenerate solitons. The results presented in this paper are expected to be useful in Bose-Einstein condensates, nonlinear optics, plasma physics, and other closely related fields where the LSRI phenomenon plays a significant role in governing the evolution of vector solitons.
\\\\
\setstretch{1.075}
\noindent{\bf Acknowledgement}\\
M.L. thanks DST-ANRF, INDIA for the award of a DST-ANRF National Science Chair\\ (NSC/2020/000029) position in which S.S. is a Research Associate.\\


\noindent{\bf Declaration of Competing Interest}\\ The authors declare that they have no known competing financial interests or personal relationships that could have appeared to influence the work reported in this paper.

\appendix
\section{$N$-Non-degenerate vector soliton solution}\label{appendix}
The explicit form of the $N$-non-degenerate soliton solution can be obtained by solving the bilinear equations of Eq. (\ref{eq.1}) \cite{hirotabook}
\begin{eqnarray}
(iD_t+D_x^2)g^{(l)}\cdot f=0, ~l=1,2,~D_xD_t f\cdot f=\sum_{l=1}^{2}\sigma_l g^{(l)}g^{(l)*},
\end{eqnarray}
which results from Eq. (\ref{eq.1}) using the transformation
\bes \begin{equation}
S^{(l)}(x,t)=\frac{g^{(l)}(x,t)}{f(x,t)},~~ l=1,2, ~~L(x,t)=2\frac{\partial ^2}{\partial x^2}\ln f(x,t).\label{A.2a}
\end{equation}
Here, $g^{(l)}$'s are complex valued functions and $f$ is a real valued function. 
The Gram determinant forms of $g^{(l)}$ and $f$ are expressed as \cite{ablowitz1999pla,Kanna2009epjst}
\bea
\hspace{-1.5cm}g^{(1)}&=&
\begin{vmatrix}
	A_{mm'} & A_{mn} & I & {\bf 0} & \phi_1 \\ 
	A_{nm} & A_{nn'}  & {\bf 0} & I &   \phi_2\\
	-I & {\bf 0} & \kappa_{mm'} & \kappa_{mn} & {\bf 0'}^T\\
	{\bf 0} & -I & \kappa_{nm} &  \kappa_{nn'} & {\bf 0'}^T\\
	{\bf 0'} & {\bf 0'}& C_1 & {\bf 0'} &0
\end{vmatrix},~f=\begin{vmatrix}
	A_{mm'} & A_{mn} & I & {\bf 0} \\ 
	A_{nm} & A_{nn'}  & {\bf 0} & I \\
	-I & {\bf 0} & \kappa_{mm'} & \kappa_{mn} \\
	{\bf 0} & -I & \kappa_{nm} &  \kappa_{nn'} 
\end{vmatrix},\label{A.2b}\\ 
g^{(2)}&=&
\begin{vmatrix}
	A_{mm'} & A_{mn} & I & {\bf 0} & \phi_1 \\ 
	A_{nm} & A_{nn'}  & {\bf 0} & I &   \phi_2\\
	-I & {\bf 0} & \kappa_{mm'} & \kappa_{mn} & {\bf 0'}^T\\
	{\bf 0} & -I & \kappa_{nm} &  \kappa_{nn'} & {\bf 0'}^T\\
	{\bf 0'} & {\bf 0'}&{\bf 0'}  & C_2 &0
\end{vmatrix}.\label{A.2c}
\end{eqnarray}
The above Grammian determinants  constitute a more general bright $N$-non-degenerate vector soliton solution of the two-component LSRI system (\ref{eq.1}).  
The various elements involving in the above determinants are defined as 
\begin{eqnarray}
&&A_{mm'}=\frac{e^{\eta_m+\eta_{m'}^*}}{(k_m+k_{m'}^*)},~ A_{mn}=\frac{e^{\eta_m+\xi_{n}^*}}{(k_m+l_{n}^*)},
A_{nn'}=\frac{e^{\xi_n+\xi_{n'}^*}}{(l_n+l_{n'}^*)}, ~A_{nm}=\frac{e^{\eta_n^*+\xi_{m}}}{(k_n^*+l_{m})},
\nonumber
\\
&&\kappa_{mm'}=\frac{\psi_m^{\dagger}\nu\psi_{m'}}{2i(k_m^2-k_{m'}^{*2})},~\kappa_{mn}=\frac{\psi_m^{\dagger}\nu\psi'_{n}}{2i(l_m^2-k_{n}^{*2})},~\kappa_{nm}=\frac{\psi_n^{'\dagger}\nu\psi_{m}}{2i(k_n^2-l_{m}^{*2})},~
\kappa_{nn'}=\frac{\psi_n^{'\dagger}\nu\psi'_{n'}}{2i(l_n^2-l_{n'}^{*2})},\nonumber\\
&&~~~~~~~~~~m,m',n,n'=1,2,...,N.\nonumber\\
&&\phi_1=\begin{pmatrix} e^{\eta_{1}} & e^{\eta_{2}}&\cdot & \cdot &\cdot & e^{\eta_{N}}\end{pmatrix}^T,~\phi_2=\begin{pmatrix} e^{\xi_{1}} & e^{\xi_{2}}& \cdot &\cdot &\cdot &e^{\xi_{N}}\end{pmatrix}^T,~\psi_j=
\begin{pmatrix} \alpha_j^{(1)} & 0\end{pmatrix}^T,\nonumber\\
&&\psi_j'=\begin{pmatrix} 0 &  \alpha_j^{(2)}\end{pmatrix}^T,~C_l=-\begin{pmatrix} \alpha_1^{(l)} &  \alpha_2^{(l)}&\cdot &\cdot &\cdot &\alpha_j^{(l)}\end{pmatrix},~j=1,2,...,N,~l=1,2. \label{A.2d}
\end{eqnarray}\ees
The other matrices  are defined below:\\ ${\bf 0'}$ is a null matrix of order $(1\times N)$, $I=\nu$ is an identity matrix of order $(N\times N)$, and ${\bf 0}$ is a $(N\times N)$ null matrix. 

 We wish to remark here that the degenerate $N$-soliton solution can be obtained from the above non-degenerate $N$-soliton solution when $k_j=l_j$, $j=1,2,...,N$.  We also point out here that one can also  obtain the non-degenerate $N$-soliton solution from the degenerate $2N$-soliton solution by imposing $2N$ number of restrictions on the complex constants, $\alpha_j^{(l)}$, $j=1,2, ...,N$, $l=1,2$. 
 \section{Asymptotic Constants appearing in Section 4 }
\begin{eqnarray}
	&&\hspace{-0.9cm}d_{11}=\frac{(ic_2)^{\frac{1}{2}}(k_1-k_2)(k_1-l_1)^{1/2}|k_2-l_1|\alpha_1^{(1)}|\alpha_1^{(2)}|\Lambda_1^{1/2}\Lambda_2^{1/2}}{2\sqrt{2}(k_1-k_2^*)(k_1+k_2^*)^2(k_2-k_2^*)(k_2+k_2^*)^2|k_2-l_1^*||k_2+l_1^*|^2(k_2-l_1^*)^{\frac{1}{2}}(l_1-l_1^*)^{\frac{1}{2}}(k_1+l_1^*)^{\frac{1}{2}}(l_1+l_1^*)},\nonumber\\
	&&\hspace{-0.9cm}d_{12}=\frac{(ic_1)^{\frac{1}{2}}c_2|k_1-k_2|(k_1-k_2)^{\frac{1}{2}}(k_1+k_2)^{\frac{1}{2}}(k_1-l_1)^{\frac{1}{2}}(k_2-l_1)|\alpha_1^{(1)}||\alpha_2^{(1)}|\alpha_2^{(2*)}\alpha_2^{(1)\frac{1}{2}}}{2\sqrt{2}(k_1-k_1^*)^{\frac{1}{2}}(k_1+k_1^*)(k_1^*-k_2)^{\frac{1}{2}}(k_1^*+k_2)(k_1+k_2^*)^{\frac{1}{2}}(k_2+k_2^*)(k_1^*+l_1)^{\frac{1}{2}}(k_2-l_1)(k_2^*+l_1)^2},\nonumber\\
	&&\hspace{-0.9cm}d_{13}=\frac{-\sqrt{c_1c_2}|k_1-k_2||k_1-l_1||k_2-l_1||\alpha_1^{(1)}||\alpha_1^{(2)}|[\sigma_1|\alpha_2^{(1)}|^2+\sigma_2|\alpha_2^{(2)}|^2]^{1/2}\Lambda_3^{1/2}}{4(k_1-k_1^*)^{\frac{1}{2}}(k_1+k_1^*)|k_1-k_2^*||k_1+k_2^*|^2(k_2-k_2^*)(k_2+k_2^*)^2|k_2-l_1^*||k_1+l_1^*||k_2+l_1^*|^2(l_1-l_1^*)^{\frac{1}{2}}(l_1+l_1^*)},\nonumber\end{eqnarray}\begin{eqnarray}
	&&\hspace{-0.9cm}d_{14}=\frac{\sqrt{c_1c_2}|k_1-k_2||k_2-l_1||\alpha_1^{(1)}|\alpha_1^{(2)}|\Lambda_4^{1/2}\Lambda_5^{1/2}}{4(k_1-k_1^*)^{\frac{1}{2}}(k_1+k_1^*)|k_1-k_2^*|(k_2-k_2^*)(k_2+k_2^*)^2|k_2-l_1^*||k_2+l_1^*|^2|k_1+k_2^*|^2(l_1-l_1^*)^{\frac{1}{2}}(l_1+l_1^*)},\nonumber\\
	&&\hspace{-0.9cm}d_{15}=-\frac{c_1c_2|k_1-k_2||k_2-l_1||\alpha_1^{(1)}||\alpha_2^{(1)}||\alpha_1^{(2)}||\alpha_2^{(2)}|}{4|k_1-k_2^*||k_1+k_2^*|^2(k_2+k_2^*)|k_1+l_1^*||k_2-l_1^*||k_2+l_1^*|^2},\nonumber\\
	&&\hspace{-0.9cm}d_{21}=\frac{c_1(ic_2)^{\frac{1}{2}}(k_1-k_2)(k_1-l_1)^{\frac{1}{2}}(k_2-l_1)^{\frac{1}{2}}(k_2+l_1)^{\frac{1}{2}}|k_2-l_1|\alpha_1^{(1)}\alpha_2^{(1)*}\alpha_2^{(2)}|\alpha_1^{(2)}|}{2\sqrt{2}(k_1-k_2^*)(k_1+k_2^*)^2(k_2+k_2^*)|k_2+l_1^*|(k_2+l_1^*)^{\frac{1}{2}}(l_1-l_1^*)^{\frac{1}{2}}(k_1+l_1^*)^{\frac{1}{2}}(k_2-l_1^*)^{\frac{1}{2}}(l_1+l_1^*)},\nonumber\\
	&&\hspace{-0.9cm}d_{22}=\frac{i(ic_1)^{\frac{1}{2}}|k_1-k_2|(k_1-l_1)^{\frac{1}{2}}(k_2-l_1)\alpha_1^{(2)}|\alpha_1^{(1)}|\Lambda_6^{1/2}\Lambda_7^{1/2}}{2\sqrt{2}(k_1-k_1^*)^{\frac{1}{2}}(k_1+k_1^*)(k_2-k_2^*)(k_2+k_2^*)^2|k_1-k_2^*||k_1+k_2^*|^2(k_2^*-l_1)(k_2^*+l_1)^2(k_1^*+l_1)^{\frac{1}{2}}},\nonumber\\
	&&\hspace{-0.9cm}\nu_{11}=\frac{-\sigma_1\sigma_2|k_1-k_2|^2|k_1-l_1||k_2-l_1|^2|\alpha_1^{(1)}|^2|\alpha_1^{(2)}|^2[\sigma_1|\alpha_1^{(1)}|^2+\sigma_2|\alpha_1^{(2)}|^2]^{1/2}\Lambda_3^{1/2}\Lambda_4^{1/2}\Lambda_5^{1/2}}{16\hat{\lambda}|k_1-k_2^*|^2(k_2-k_2^*)^2(k_2+k_2^*)^4|k_2-l_1^*|^2|k_1+l_1^*||k_2+l_1^*|^4|k_1+k_2^*|^4}\nonumber\\
	&&\hspace{-0.9cm}\nu_{12}=\frac{\sigma_1\sigma_2\sqrt{\sigma_1\sigma_2}|k_1-k_2|^2|k_2-l_1|^2|\alpha_1^{(1)}|^2|\alpha_1^{(2)}|^2|\alpha_2^{(1)}||\alpha_2^{(2)}|\Lambda_4^{1/2}\Lambda_5^{1/2}}{16\sqrt{\hat{\lambda}}|k_1-k_2^*|^2(k_2-k_2^*)(k_2+k_2^*)^3|k_1+k_2^*|^4|k_2-l_1^*|^2|k_2+l_1^*|^4|k_1+l_1^*|},\nonumber\\
	&&\hspace{-0.9cm}\nu_{13}=\frac{\sigma_1\sigma_2\sqrt{\sigma_1\sigma_2}|k_1-k_2|^2|k_2-l_1|^2|k_1-l_1||\alpha_1^{(1)}|^2|\alpha_1^{(2)}|^2|\alpha_2^{(1)}||\alpha_2^{(2)}|[\sigma_1|\alpha_2^{(1)}|^2+\sigma_2|\alpha_2^{(2)}|^2]^{\frac{1}{2}}\Lambda_3^{\frac{1}{2}}}{16\sqrt{\hat{\lambda}}|k_1-k_2^*|^2|k_1+k_2^*|^4(k_2-k_2^*)(k_2+k_2^*)^3|k_2-l_1^*|^2|k_1+l_1^*|^2|k_2+l_1^*|^4},\nonumber\\
	&&\hspace{-0.9cm}\nu_{14}=(k_{1R}-l_{1R})^2e^{\hat{\delta}_2+\hat{\delta}_3}+(k_{1R}+l_{1R})^2e^{\hat{\delta}_1+\hat{\delta}_4},\nonumber\\
	&&\hspace{-0.9cm}\Phi_1^-=\frac{1}{2}\ln \frac{\sigma_2(k_1-l_1)|k_2-l_1|^2|\alpha_1^{(2)}|^2\Lambda_2}{2i\Lambda_1|k_2-l_1^*|^2|k_2+l_1^*|^4(k_2-l_1^*)(l_1+l_1^*)^2(k_1+l_1^*)(l_1-l_1^*)},\nonumber\\
	&&\hspace{-0.9cm}\Phi_2^-=\frac{1}{2}\ln\frac{\sigma_1(k_1-k_2)^2(k_1+k_2)(k_1^*-k_2^*)(k_1-l_1)|\alpha_1^{(1)}|^2\alpha_2^{(1)*}}{2i(k_1-k_1^*)(k_1+k_1^*)^2(k_1^*-k_2)(k_1^*+k_2)^2(k_1+k_2^*)(k_1^*+l_1)},\nonumber\\
	&&\hspace{-0.9cm}\Phi_3^-=\frac{1}{2}\ln\frac{-\sigma_1\sigma_2|k_1-k_2|^2|k_1-l_1|^2|k_2-l_1|^2|\alpha_1^{(1)}|^2|\alpha_1^{(2)}|^2\Lambda_3^{1/2}}{4\sqrt{\hat{\lambda}}|k_1-k_2^*|^2|k_1+k_2^*|^4|k_2-l_1^*|^2|k_1+l_1^*|^2|k_2+l_1^*|^4[\sigma_1|\alpha_1^{(1)}|^2+\sigma_2|\alpha_1^{(2)}|^2]},\nonumber\\
	&&\hspace{-0.9cm}\Phi_4^-=\frac{1}{2}\ln\frac{\sigma_1|k_1-k_2|^2|k_2-l_1^*|^2|k_2+l_1^*|^4(l_1-l_1^*)(l_1+l_1^*)^2\Lambda_5|\alpha_1^{(1)}|^2}{\sigma_2|k_2-l_1|^2(k_1-k_1^*)(k_1+k_1^*)^2|k_1-k_2^*|^2|k_1+k_2^*|^4|\alpha_1^{(2)}|^2\Lambda_4},\nonumber\\
	&&\hspace{-0.9cm}\Phi_5^-=\frac{1}{2}\ln\frac{(k_1-k_2)(k_1^*-k_2)(k_1^*+k_2)(k_2^*-l_1^*)(k_1^*+l_1)(k_2^*-l_1)(k_2^*+l_1)\alpha_1^{(1)}\alpha_1^{(2)*}\alpha_2^{(1)*}\alpha_2^{(2)}}{(k_1^*-k_2^*)(k_1-k_2^*)(k_1+k_2^*)(k_2-l_1)(k_1+l_1^*)(k_2-l_1^*)(k_2+l_1^*)\alpha_1^{(1)*}\alpha_1^{(2)}\alpha_2^{(1)}\alpha_2^{(2)*}},\nonumber\\
	&&\hspace{-0.9cm}\Phi_6^-=\frac{1}{2}\ln\frac{\sigma_2(k_1-l_1)(k_2-l_1)|k_2-l_1|^2(k_2+l_1)|\alpha_1^{(2)}|^2}{2i|k_2+l_1^*|^2(k_2+l_1^*)(k_1+l_1^*)(k_2-l_1^*)(l_1-l_1^*)(l_1+l_1^*)^2},\nonumber\\
	&&\hspace{-0.9cm}\Phi_7^-=\frac{1}{2}\ln\frac{-\sigma_1|k_1-k_2|^2(k_1-l_1)|\alpha_1^{(1)}|^2\Lambda_7}{2i\Lambda_6(k_1-k_1^*)(k_1+k_1^*)^2|k_1-k_2^*|^2|k_1+k_2^*|^4(k_1^*+l_1)},
\end{eqnarray}
where 
\bea
&&\hspace{-0.9cm}\Lambda_1=[\sigma_1|\alpha_2^{(1)}|^2(k_1^2-k_2^2)+\sigma_2|\alpha_2^{(2)}|^2(k_1^2-k_2^{*2})],~\hat{\lambda}=(k_1-k_1^*)(k_1+k_1^*)^2(l_1-l_1^*)(l_1+l_1^*)^2,\nonumber\\
&&\hspace{-0.9cm}\Lambda_2=[\sigma_1(k_1^2-k_2^2)|k_2+l_1^*|^2|k_2-l_1^*|^2|\alpha_2^{(1)}|^2+\sigma_2(k_1^2-k_2^{*2})|k_2+l_1|^2|k_2-l_1|^2|\alpha_2^{(2)}|^2],\nonumber\\
&&\hspace{-0.9cm}\Lambda_3=[\sigma_1|k_1-k_2|^2|k_1+k_2|^2|k_2-l_1^*|^2|k_2+l_1^*|^2|\alpha_2^{(1)}|^2+\sigma_2|k_1-k_2^*|^2|k_1+k_2^*|^2|k_2-l_1|^2|k_2+l_1|^2|\alpha_2^{(2)}|^2],\nonumber\\
&&\hspace{-0.9cm}\Lambda_4=[\sigma_1|k_2-l_1^*|^2|k_2+l_1^*|^2|\alpha_2^{(1)}|^2+\sigma_2|k_2-l_1|^2|k_2+l_1|^2|\alpha_2^{(2)}|^2],\nonumber\\
&&\hspace{-0.9cm}\Lambda_5=[\sigma_1|k_1-k_2|^2|k_1+k_2|^2|\alpha_2^{(1)}|^2+\sigma_2|k_1-k_2^*|^2|k_1+k_2^*|^2|\alpha_2^{(2)}|^2],\nonumber\\
&&\hspace{-0.9cm}\Lambda_6=[\sigma_1(k_2^{*2}-l_1^2)|\alpha_2^{(1)}|^2+\sigma_2(k_2^{2}-l_1^2)|\alpha_2^{(2)}|^2],\nonumber\\
&&\hspace{-0.9cm}\Lambda_7=[\sigma_1(k_2^{*2}-l_1^2)|k_1-k_2|^2|k_1+k_2|^2|\alpha_2^{(1)}|^2+\sigma_2|k_1-k_2^*|^2|k_1+k_2^*|^2(k_2^{2}-l_1^2)|\alpha_2^{(2)}|^2].
\eea
\setstretch{01.20}

\end{document}